\documentclass[twocolumn,english,prb,aps,superscriptaddress]{revtex4-1}
\usepackage{amsmath}
\usepackage{amssymb}
\usepackage{graphicx}
\usepackage[USenglish]{babel}
\usepackage[titletoc]{appendix}
\usepackage{chngcntr}

\DeclareMathOperator{\sgn}{sgn}
\newcommand{\beq}{\begin{equation}}
\newcommand{\eeq}{\end{equation}}
\newcommand{\bea}{\begin{eqnarray}}
\newcommand{\eea}{\end{eqnarray}}

\renewcommand{\[}{\left[}
\renewcommand{\]}{\right]}

\begin{document}

\title{Raman resonance in iron-based superconductors: The magnetic scenario}

\author{Alberto Hinojosa}
\affiliation{Department of Physics, University of Minnesota,
Minneapolis, Minnesota 55455, USA}
\author{Jiashen Cai}
\affiliation{Department of Physics, University of Minnesota,
Minneapolis, Minnesota 55455, USA}
\author{Andrey V. Chubukov}
\affiliation{Department of Physics, University of Minnesota,
Minneapolis, Minnesota 55455, USA}

\begin{abstract}
We perform  theoretical analysis of polarization-sensitive Raman spectroscopy on NaFe$_{1-x}$Co$_x$As, EuFe$2$As$_2$, SrFe$_2$As$_2$,  and Ba(Fe$_{1-x}$Co$_x$)$_2$As$_2$, focusing on two features seen in the $B_{1g}$ symmetry channel (in one Fe unit cell notation): the strong temperature dependence of the static, uniform Raman response in the normal state and the existence of a collective mode in the superconducting state. We show that both features can be explained by the coupling of fermions to pairs of magnetic fluctuations via the Aslamazov-Larkin process.  We first analyze magnetically-mediated Raman intensity at the leading two-loop order  and then include interactions between pairs of magnetic fluctuations.  We show that   the full Raman intensity in the $B_{1g}$ channel can be viewed as the result of the coupling of light to Ising-nematic susceptibility via  Aslamazov-Larkin process. We argue that the singular temperature dependence in the normal state is the combination of the temperature dependencies of the Aslamazov-Larkin vertex and of Ising-nematic susceptibility. We discuss two scenarios for the resonance below $T_c$.  In one, the resonance is due to the development of a pole in the fully renormalized Ising-nematic susceptibility. The other is the orbital excitonic scenario, in which spin fluctuations generate an attractive interaction between low-energy fermions.
\end{abstract}

\maketitle

\section{Introduction}
Raman scattering in Fe-based superconductors has attracted  substantial interest in the past few years due to the number of new features associated with multi-orbital/multi-band nature of these materials (see, e.g., Refs.  [\onlinecite{boyd,balen,cek,blum_scr,dev_scal,sugai,mazin,wen}]).
 The subject of this paper is the theoretical analysis of the features in Raman scattering revealed by polarization-sensitive Raman spectroscopy in the normal and the superconducting states of Fe-based materials NaFe$_{1-x}$Co$_x$As [\onlinecite{Thorsmolle}], $A$Fe$_2$As$_2$, $A=$Eu,Sr  [\onlinecite{Girsh_b,Yang}], and Ba(Fe$_{1-x}$Co$_x$)$_2$As$_2$ [\onlinecite{BaFeCoAs1,BaFeCoAs2}]. Polarized light was used to probe the Raman response in different symmetry channels, classified by the irreducible representations of the $D_{4h}$ crystallographic point group [\onlinecite{boyd}].  In the normal state  the real part of the (almost) static and uniform Raman susceptibility in the $B_{1g}$ channel in one-iron unit cell notation
  (same as the $B_{2g}$ channel in two-iron notation used in Refs. [\onlinecite{Thorsmolle,Girsh_b,BaFeCoAs1}])
    is strongly temperature dependent---it increases below 300 K roughly as $1/(T-T_0)$, where $T_0$
 is positive at small doping, but changes sign and becomes negative above optimal doping.
 In the superconducting state  the imaginary part of the $B_{1g}$ Raman susceptibility  displays a strong resonance-type peak at around 50 cm$^{-1}$. There is no such resonance peak in other channels, although the Raman intensity in the $A_{1g}$ channel does show a broad maximum at a somewhat higher frequency [\onlinecite{Thorsmolle}].

In the effective mass approximation (in which
the coupling of light to fermions is proportional to the square of vector potential) the measured Raman intensity in a particular channel
($A_{1g}$, $B_{1g}$, $B_{2g}$,$\ldots$) is proportional to the imaginary part of the fully renormalized particle-hole polarization bubble $\chi_R (\mathbf{p}, \Omega)$ with proper symmetry factors in the vertices, taken at vanishingly small transferred momentum $\mathbf{p}$ and finite transferred frequency $\Omega$ [\onlinecite{klein_1, dever, cordona}].

The free-fermion polarization bubble vanishes in the normal state for $\Omega > v_F p$, where $v_F$ is the Fermi velocity,
 and obviously it cannot account for the observed strong temperature dependence of $B_{1g}$ Raman intensity above $T_c$.
 It is nonzero in the superconducting state, but does not display a peak.
   The effect  must then come from
the renormalization of the Raman vertex due to coupling to some low-energy fluctuations (final state interaction in Raman literature~[\onlinecite{klein}]). This coupling may come from three different sources. First, the $B_{1g}$ Raman vertex changes sign under $k_x \leftrightarrow k_y$ (i.e., under interchanging the $x$ and $y$ directions in real space), hence it couples to strain (structural fluctuations). Second,
 the $B_{1g}$  vertex is anti-symmetric with respect to the interchange between the
iron
$d_{xz}$ and $d_{yz}$ orbitals and hence couples to orbital fluctuations. Third, symmetry allows the coupling between the $B_{1g}$ Raman vertex and Ising-nematic spin fluctuations [the ones that distinguish between the magnitudes of spin-density-wave order parameters with
ordering vectors
$(0,\pi)$ and $(\pi,0)$] because both are anti-symmetric with respect to 90$^\circ$ rotations in the momentum space.

Structural fluctuations, orbital fluctuations, and Ising-nematic spin-fluctuations are the three key candidates to drive the nematic order, observed in most of the Fe-based materials.  How to choose the primary
 candidate among these three
has become
one of the key issues in the studies of Fe-based superconductors [\onlinecite{review}]. We intend to verify whether the Raman data can help distinguish between the three candidates.

The effects of structural and orbital fluctuations has been discussed before (see Refs. [\onlinecite{GP}], [\onlinecite{yama}] and references therein).
  Structural fluctuations (acoustic phonons associated with strain) practically do not affect the Raman intensity because the coupling to phonons changes the $B_{1g}$ Raman susceptibility $\chi_{R}(\mathbf{p}, \Omega)$  to
 \begin{equation}
{\tilde \chi}_R(\mathbf{p}, \Omega)=\left[\chi_{R}^{-1}(\mathbf{p}, \Omega)-\frac{\lambda^2_{ph} p^2}{C^2_{ph} p^2-\Omega^2}\right]^{-1},
\label{zz_2}
\end{equation}
where  $\lambda_{ph}$ is electron-phonon coupling and  $C_{ph}$ is the elastic constant for orthorhombic strain.
Such coupling is relevant in the static limit, where ${\tilde \chi}^{-1}_R(\mathbf{p},0) = \chi_{R}^{-1}(\mathbf{p},0) - (\lambda_{ph}/C_{ph})^2$
  differs from  $\chi_{R}^{-1}(\mathbf{p},0)$ by a constant term,
  but is irrelevant in the limit of vanishing $p$ and finite $\Omega$, where Raman measurements have been performed [\onlinecite{Thorsmolle,BaFeCoAs1,BaFeCoAs2}].
(In the $B_{1g}$ channel the minimum $p$ is, strictly speaking, nonzero [\onlinecite{blumberg}], but is generally of order of inverse system size).

 Orbital fluctuations do affect the $B_{1g}$ Raman susceptibility via renormalizations involving particular combinations of intra-orbital and inter-orbital Hubbard and Hund terms,
  compatible with the fact that the $B_{1g}$ Raman vertex changes sign between $d_{xz}$  and $d_{yz}$ orbitals.
  By orbital fluctuations we mean fluctuations which renormalize $B_{1g}$ Raman vertex by inserting series of ladder and bubble diagrams, as shown schematically in
   Fig. \ref{fig:ladders} and in more detail in Figs. \ref{fig:v_renorm_h} and \ref{fig:v_renorm_e} in the Appendix.

 \begin{figure}[htb]
	\centering
		\includegraphics[width=0.45\textwidth]{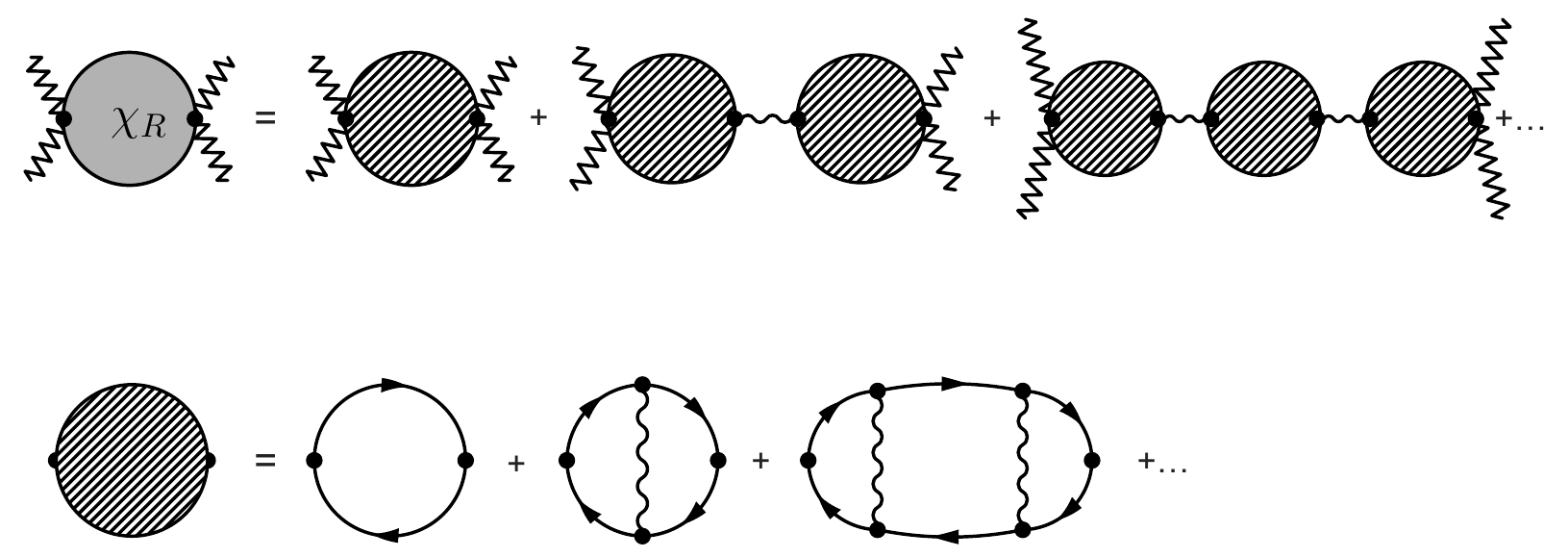}
	\caption{Ladder and bubble diagrams for the renormalization of the $B_{1g}$ Raman intensity within RPA. Each combination of Green's functions with equal momenta gives
 a particle-hole polarization bubble $\Pi_{B_{1g}} (\Omega)$ and the combination of vertical and horizontal interaction lines gives the coupling $\lambda$. We obtain $\lambda$ explicitly in the Appendix for a system with dominant intra-orbital Hubbard interaction.}
	\label{fig:ladders}
\end{figure}

 In the band basis, the
   interaction lines in these diagrams are
    Hubbard and Hund terms dressed by
   coherence factors, associated with the transformation from the orbital to the band basis, and projected into $B_{1g}$ channel.
   (Each interaction contains one incoming and one outgoing fermion with momentum ${\bf k}$  and one incoming
   and one outgoing fermion with momentum ${\bf p}$, and its $B_{1g}$ component is proportional to $\cos{2\theta_k} \cos{2\theta_p}$, where $\tan{\theta_k} = k_y/k_x$).
     In the approximation when only ladder and bubble diagrams are kept [often called
    random-phase approximation (RPA)], the
 Raman response in the $B_{1g}$ channel is given by (see, e.g., Ref. [\onlinecite{Gallais_1}]),
\begin{equation}{\label{eq:Raman-RPA}}
	\chi_R(\Omega)=\frac{\Pi_{B_{1g}}(\Omega)}{1+\lambda \Pi_{B_{1g}}(\Omega)},
\end{equation}
where $\chi_R (\Omega) = \chi_R (p \to 0, \Omega)$, $\lambda$ is the proper combination  of interactions in $B_{1g}$ channel, and $\Pi_{B_{1g}} (\Omega)$ is the particle-hole polarization function at zero momentum transfer, summed over all bands with the $B_{1g}$ form factor.
[We define $\Pi_{B_{1g}} (\Omega)$ as 
$i\int \mathrm{d}^2k \mathrm{d} \nu /(2\pi)^3 \times G(k, \nu) G(k, \nu + \Omega)$.
With this sign convention, Re $\Pi_{B_{1g}} (\Omega)$ in a superconductor is positive at $\Omega < 2\Delta$,
 where $\Delta$ is the superconducting gap].
In the language of Fermi-liquid theory, $\lambda \Pi_{B_{1g}}(\Omega=0)$ is the $B_{1g}$ component of the quasiparticle interaction function, and $ \lambda\Pi_{B_{1g}}(\Omega=0)=-1$ would correspond to $B_{1g}$ Pomeranchuk instability.

  Because free-fermion $\Pi_{B_{1g}}(\Omega)$ vanishes in the normal state (the poles of both Green's functions are in the same frequency half plane),  Eq. (\ref{eq:Raman-RPA}) cannot explain
  the normal state behavior of the Raman response.  However, $\chi_R(\Omega)$ from Eq. (\ref{eq:Raman-RPA}) is nonzero in a superconductor, because $\Pi_{B_{1g}} (\Omega)$ becomes nonzero, and for negative $\lambda$ it
   displays a resonance peak.
     The resonance develops by the same reason as the excitonic spin resonance in a $d$-wave superconductor [\onlinecite{neutron_res}]: the imaginary part of the polarization function  $\Pi (\Omega) $ vanishes for $\Omega<2\Delta$,
while the  real part of $\Pi (\Omega) $ is positive and diverges at $\Omega=2\Delta$.   As the result, for negative $\lambda$, the denominator in (\ref{eq:Raman-RPA}) is guaranteed to pass through zero at some frequency below $2\Delta$,
  and a sharp resonance in $\chi_R (\Omega)$ appears at this frequency [\onlinecite{Gallais_1}].

  This would be the most natural explanation of the Raman resonance. The problem, however, is how to
	justify that $\lambda$  is negative, i.e., that there is
	an attraction in the $B_{1g}$ ($d$-wave) charge Pomeranchuk channel.
   If intra-orbital Hubbard repulsion is the dominant interaction term, $\lambda$ is  definitely positive and orbital fluctuations do not give rise to the resonance in the Raman intensity  (we show this in the Appendix).
     The coupling $\lambda$ does become negative when inter-orbital interaction $U'$ and exchange Hund interaction $J$ are included and $U'$ is set to be
      about equal to $U$ and  larger than $J$ [\onlinecite{kontani}]. However, the relation $U' \approx U$ gets broken once one starts integrating out high-energy fermionic excitations [\onlinecite{chu_1}] or includes lattice effects  [\onlinecite{scalapino_1}]. In a generic case it is natural to expect that the intra-orbital Hubbard interaction is the strongest interaction between Fe $d$-orbitals. If so,
     the coupling $\lambda$ is positive and there is no resonance in $\chi_R (\Omega)$ within RPA.

 In this paper we analyze whether the increase of $\chi_R (\Omega)$ in the $B_{1g}$ channel in the normal state and the  sharp peak in  the Raman response in this channel below $T_c$ can be due to Ising-nematic spin fluctuations associated with stripe magnetism. The advantage of the magnetic scenario  is that Ising-nematic fluctuations are enhanced even when intra-orbital Hubbard interaction is the dominant interaction between fermions. The only requirement
is that the magnetic order should be stripe rather than checkerboard [\onlinecite{FC}].

\begin{figure}[htb]
	\centering
		\includegraphics[width=0.45\textwidth]{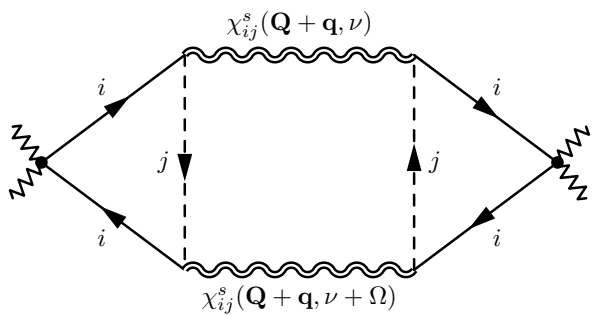}
	\caption{The lowest order (two-loop) AL diagram for the Raman intensity.  The momenta $\mathbf{Q}_1 = (\pi,0)$ and $\mathbf{Q}_2 = (0,\pi)$.
The solid and dashed lines represent fermions from different bands
	with band indices $i$ and $j$.
	The sinuous lines represent spin fluctuations and the external jagged lines are the coupling to photons.}
	\label{fig:AL_diag}
\end{figure}

  The coupling of the Raman vertex to a pair of spin fluctuations with momenta near $\mathbf{Q} = (0,\pi)$ [or $(\pi,0)$] occurs via the Aslamazov-Larkin (AL) process.
   The corresponding diagram is  presented in Fig. \ref{fig:AL_diag}.  AL diagrams for Raman scattering have been earlier discussed in Ref. [\onlinecite{AL}], but in a different
context.
 The lowest-order AL type diagram (the one shown in Fig. \ref{fig:AL_diag}) contains two triangular vertices made out of fermionic Green's functions from hole and electron bands, and two spin-fluctuation propagators. The vertex between fermions and spin fluctuations can be obtained by
 decomposing the antisymmetrized interaction into spin and charge parts, by
 focusing on spin-spin part, and by using the Hubbard-Stratonovich method to transform the interaction between fermionic spins into spin-spin interaction between  a fermion
   and  a collective bosonic variable in the spin channel [\onlinecite{FC}].

     We show that in the normal state, above a certain temperature,
 each  triangular vertex $\Gamma_{tr}$ scales as $1/T$, while the convolution of the two  spin propagators at equal frequencies
  (i.e., $T \sum_{\nu_n} \int d^2 \mathbf{q}/(2\pi)^2\times [\chi^s (\mathbf{Q}+\mathbf{q},\nu_n)]^2$) scales as $T$. As the consequence, the Raman susceptibility from Fig. \ref{fig:AL_diag} scales as $1/T$.
   This holds in both $A_{1g}$ and $B_{1g}$ channels.  Higher-order processes, shown in Fig. \ref{fig:Box} and in more detail in Fig. \ref{fig:higher_order} below, however, distinguish between $A_{1g}$ and $B_{1g}$ Raman responses.
    Namely, an attractive interaction between magnetic fluctuations in the $B_{1g}$ channel increases $B_{1g}$ Raman response and changes
   $1/T$ dependence into $1/(T-T_0)$ (see Refs. [\onlinecite{Raman_nematic}, \onlinecite{Raman_nematic_khodas}] and Sec.III below), while the (much stronger) repulsive interaction in the $A_{1g}$ channel almost completely eliminates the temperature dependence of $A_{1g}$ Raman response.  This behavior fully agrees with the data.

In the superconducting state,
the $1/T$ behavior of $\Gamma_{tr}$ is cut by the gap opening, while $\chi^2 (\Omega)= \int d\nu d^2 \mathbf{q}/(2\pi)^3 \times \chi^s (\mathbf{q}+\mathbf{Q}, \nu) \chi^s (\mathbf{q}+\mathbf{Q}, \nu + \Omega)$ becomes singular.
The  real part of $\chi^2 (\Omega)$ diverges at $\Omega=2\Omega_{mag}$, where $\Omega_{mag}$ is the minimal frequency of the magnetic resonance in the superconducting state, and its imaginary part jumps at this frequency from zero to a finite value.
 Higher-order terms change $\chi^2 (\Omega)$ into $\chi_{I-nem} (\Omega) = \chi^2 (\Omega)/[1 + 2 g \chi^2 (\Omega)]$ (see Sec. III),  where $g$ is negative (attractive) when magnetic order is of stripe type [\onlinecite{FC}].
 Approximating the triangular vertex $\Gamma_{tr}$ by a constant we then obtain
 \begin{equation}\label{eq:RPA_AL}
	\chi_R(\Omega)=\Gamma^2_{tr} \chi_{I-nem} (\Omega) = \Gamma^2_{tr}~\frac{\chi^2(\Omega)}{1  + 2 g \chi^2(\Omega)}.
\end{equation}
The combination of $g <0$ and the fact that the real part of $\chi^2(\Omega)$ is positive and diverges at $\Omega=2\Omega_{mag}$  guarantees that
$1  + 2 g \chi^2(\Omega)$ passes through zero  at some $\Omega = \Omega_{res,1} <2\Omega_{mag}$. At this frequency the Raman intensity Im $\chi_R (\Omega)$  displays a $\delta$-functional peak.

\begin{figure}[htb]
	\centering
		\includegraphics[width=0.45\textwidth]{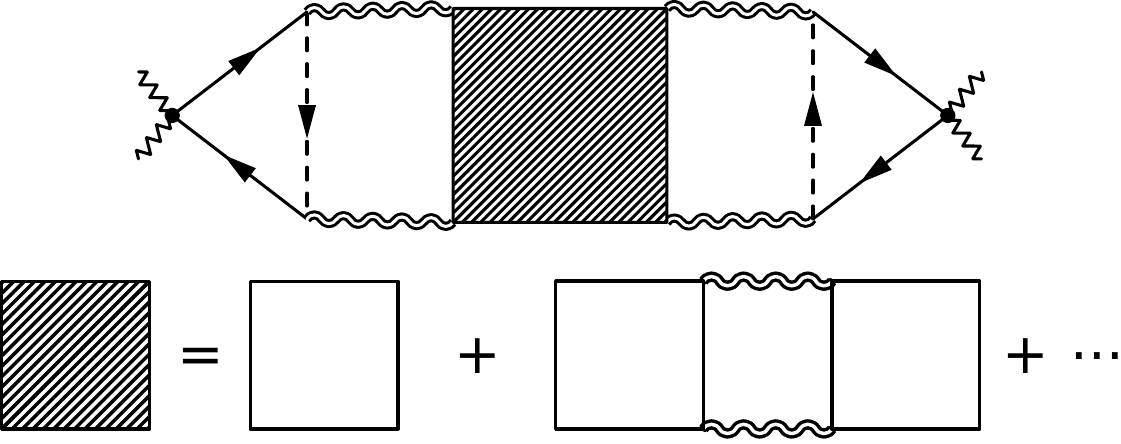}
	\caption{Schematic representation of higher-order contributions to the Raman response. We show higher-order terms in more detail in Fig. \protect\ref{fig:higher_order}.}
	\label{fig:Box}
\end{figure}

We show that approximating $\Gamma_{tr}$ by a constant is justified if typical fermionic frequencies in the triangular diagram for $\Gamma_{tr}$ are larger than $\Omega_{mag}$.
 These relevant frequencies are of order $\Delta$, hence the analysis is justified when $\Omega_{mag} \ll \Delta$.  This holds if the inverse magnetic correlation length $m_s$ is small enough because
 $\Omega_{mag} \propto m_s$
 [\onlinecite{ac_1}].  In the Fe-based materials, in which $B_{1g}$ resonance has been observed, the situation is, however,  somewhat different: neutron scattering data for NaFe$_{1-x}$Co$_x$As with $x=0.045$ show [\onlinecite{neutron_0.045}] that $\Omega_{mag}\approx 7$ meV, while $\Delta = 5-5.5$ meV [\onlinecite{prx}],
 i.e., $\Delta$ is somewhat smaller than $\Omega_{mag}$. Similarly, for Ba(Fe$_{1-x}$Co$_x$)$_2$As$_2$ with x=0.075, $\Omega_{mag} \approx 9.5$ meV [\onlinecite{Inosov}], while $\Delta=4.5-5$meV on the electron pocket  and 7 meV on the hole pocket [\onlinecite{Tera}], so again $\Delta<\Omega_{mag}$. In this situation, there is no good reason to treat $\Gamma_{tr}$ as a constant, independent on $\Omega_{mag}$, because two fermionic frequencies in the triangular loop differ by $\Omega_{mag}$.

\begin{figure}[htb]
	\centering
		\includegraphics[width=0.45\textwidth]{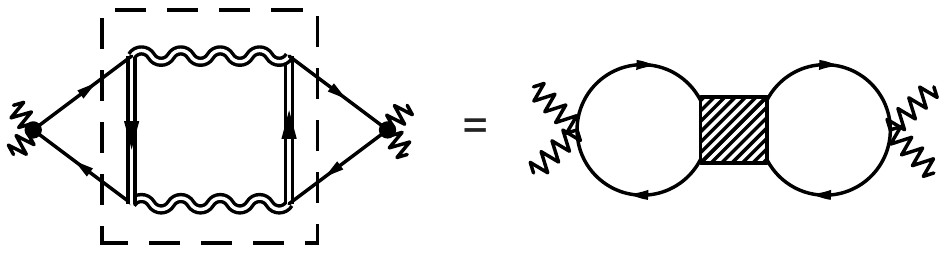}
	\caption{Interpretation of the two-loop AL diagram as consisting of two particle-hole bubbles with zero momentum transfer (unshaded circles), separated by the magnetically-mediated four-fermion interaction. The notations are the same as in Fig. \protect \ref{fig:AL_diag}. Higher-order terms are shown in Fig. \protect\ref{fig:reinter}.}
	\label{fig:reinter_1}
\end{figure}

In view of this complication, we also analyze another scenario for the $B_{1g}$ resonance below $T_c$, namely, that
 the resonance
 originates from the $2\Delta$ singularity of the particle-hole polarization $\Pi_{B_{1g}} (\Omega)$ in an $s$-wave superconductor,
 like in the pure orbital fluctuation scenario, while spin fluctuations renormalize the original, likely repulsive, coupling
  $\lambda$ in Eq. (\ref{eq:Raman-RPA}) into the effective coupling $\lambda_{eff}$ (see Figs. \ref{fig:reinter_1}).
  This scenario  is justified in the opposite limit when $\Omega_{mag}$ is assumed to be
  much larger than $\Delta$. The strong inequality never holds because $\Omega_{mag} < 2\Delta$, but weak inequality may be already enough number-wise.
   We show that
   $\lambda_{eff}$  turns out to be negative
    (i.e., attractive) near a nematic instability.
     For negative $\lambda_{eff}$,  $1 +\lambda_{eff} \Pi_{B_{1g}} (\Omega)$ necessary vanishes
    at some frequency $\Omega_{res,2}$ below $2\Delta$, where $\mathrm{Im}  \Pi_{B_{1g}} (\Omega)$ is zero, and this leads to an excitonic-type resonance in the $B_{1g}$ Raman response.

     Such a scenario has been proposed in earlier works [\onlinecite{BaFeCoAs2,Gallais_1}] based on the phenomenological argument that Ising-nematic and
     orbital order parameters break the same $C_4$ symmetry and hence are linearly  coupled in the Landau functional (see e.g., Ref. [\onlinecite{RJ}]). A bilinear term with a constant
     prefactor $A$  was argued to give rise to the renormalization of $\lambda$ into $\lambda_{eff} = \lambda - A^2 \chi_{I-nem}$. The latter is obviously negative when $\chi_{I-nem}$ is large.
   We compute the renormalization of $\lambda$ within our microscopic model. We show that $\lambda_{eff}$ does become negative and scales as $\chi_{I-nem}$.
      However, the prefactor  is not $A^2$ and is nonzero only if one includes
    the non-analytic dynamical Landau damping term into the spin propagator.  If spin-fluctuation propagator is approximated by its static part,
     $\lambda_{eff} =0$.   We explain the difference between the  coupling  between Pomeranchuk order parameter and two spin fluctuations (= A) and between Ising-nematic and orbital (Pomeranchuk) order parameters, which actually gives rise to the renormalization of $\lambda$.

The outcome of this study is that the resonance in $B_{1g}$ Raman intensity is of purely magnetic origin at $\Omega_{mag} \ll \Delta$---it is due to the pole in $\chi_{I-nem} (\Omega)$ at $\Omega_{res,1}$.  At $\Omega_{mag} > \Delta$ the resonance of fermionic origin---it develops by the same reason as in purely orbital scenario, due to the emergence of the excitonic pole in the ladder series of particle-hole bubbles with $B_{1g}$ vertices. However, the attraction between fermions in the $B_{1g}$ channel comes from magnetically mediated interaction.

 In FeSCs, $\Omega_{mag}$ and $\Delta$ are  comparable, in which case the actual resonance is likely the mixture of the  nematic and the excitonic resonances.
    In NaFe$_{1-x}$Co$_x$As with $x=0.045$,  the $B_{1g}$ peak is seen at 7.1 meV [\onlinecite{Thorsmolle}],
      which is below both $2\Delta$ and $2\Omega_{mag}$. Similarly, in Ba(Fe$_{1-x}$Co$_x$)$_2$As$_2$ with x=0.061, the B1g peak is seen at 8.7 meV [\onlinecite{BaFeCoAs1}] , which is also below both $2 \Delta$ and $2\Omega_{mag}$.


The paper is organized as follows: In  Sec. \ref{sec:Raman} we evaluate analytically the two-loop AL-type diagram for $B_{1g}$ Raman scattering.
 In Sec. \ref{sec:T_dep} we show that in the normal state this contribution to $\chi_R (\Omega)$ is strongly temperature dependent. The temperature dependence is roughly $1/T$.   We argue that higher-order terms, which include interactions between pairs of spin fluctuations, replace $1/T$ dependence into more singular $1/(T-T_0)$.
 In Sec.  \ref{sec:Enhancement} we extend the analysis to the superconducting state.
  In Sec. \ref{sec:Raman_n} we
  argue that spin excitations  with momenta near $(0,\pi$) and $(\pi,0)$ evolve below $T_c$ and become magnon-like, with minimal energy $\Omega_{mag}$.  We compute the two-loop
AL diagram assuming that the vertices that couple light to spin fluctuations saturate below $T_c$, and show that this contribution to Raman intensity becomes logarithmically singular at $\Omega =2\Omega_{mag}$. We further  show that higher-order terms, which include interactions between spin fluctuations, convert logarithmical singularity at $2\Omega_{mag}$ into a true resonance peak at an energy $\Omega_{res,1} < 2 \Omega_{mag}$.
 In Sec. \ref{sec:Raman_3} we re-interpret the two-loop AL diagram for $B_{1g}$ Raman scattering differently, as the contribution from two particle-hole polarization bubbles with an effective interaction mediated by spin fluctuations. We compute the magnetically mediated four-fermion interaction $\lambda_{mag}$  and show that it is attractive.  We argue that  higher-order terms give rise to an excitonic peak in $\chi_R (\Omega)$ at $\Omega_{res,1}$ below $2\Delta$.   In Sec. \ref{sec:comp} we discuss the interplay between this peak and the one coming from fully renormalized nematic susceptibility.   In Sec. \ref{sec:Raman_2} we present the results of numerical computation of  spin-fluctuation contribution to Raman intensity at two-loop and higher orders.
 In Sec. \ref{sec:Symmetry} we compare AL vertices for the coupling to spin fluctuations in different symmetry channels
 and show that in the $B_{2g}$ channel (in the 1-Fe zone) the vertex for the coupling of light to spin fluctuations vanishes by symmetry.
  The vertex in $A_{1g}$  channel does not vanish and is of the same order as the vertex in $B_{1g}$ channel.  We show, however, that there is no resonance in $A_{1g}$ because the interaction between spin fluctuations in this channel is strongly repulsive instead of attractive.
  We present our conclusions in Sec. \ref{sec:Conclusions}.

Throughout the paper we will be using band formalism and will be working in the 1-Fe Brillouin zone (BZ).

\section{Raman response from spin fluctuations}
\label{sec:Raman}

We consider the four-band model of NaFe$_{1-x}$Co$_x$As
and Ba(Fe$_{1-x}$Co$_x$)$_2$As$_2$
 with two hole pockets centered at  $k_x = k_y =0$ and two electron pockets centered at $(0,\pi)$ and $(\pi,0)$ in the 1 Fe Brillouin zone (BZ),
 see Refs. [\onlinecite{ARPES_NaFeAs,ARPES_NaFeCo0.0175As,ARPES_NaFeCo0.028As,ARPES_NaFeCo0.05As}].  Excitations near the hole pockets are composed out of $d_{xz}$ and $d_{yz}$ orbitals and there is 90$^o$  rotation of the orbital content near one Fermi surface compared to the other.   Excitations near electron pockets are predominantly composed out of $d_{xy}$ and $d_{xz}$ orbitals for the $(0,\pi)$ pocket and out of $d_{xy}$ and $d_{yz}$ orbitals for the $(\pi,0)$ pocket [\onlinecite{scalapino}].
  We do not include into consideration the third hole pocket, centered at $(\pi,\pi)$ in the 1Fe zone, as it is made out of $C_4$-symmetric $d_{xy}$ orbital and
   does not play any significant  role in the analysis of Raman scattering, particularly in $B_{1g}$ geometry.

The Raman response function can be calculated as a time-ordered average of density operators weighted with Raman form factors:
\begin{equation}
	\chi_R(\mathbf{p},\Omega) = -i\int \mathrm{d}t e^{i\Omega t}\left\langle T \rho_\mathbf{p}(t)\rho_\mathbf{-p}(0) \right\rangle,
\end{equation}
where
\begin{equation}
	\rho_\mathbf{p} \equiv \sum_{i,\mathbf{k},\sigma} \gamma_i(\mathbf{k}) c^\dagger_{i,\mathbf{k}+\mathbf{p},\sigma}c_{i,\mathbf{k},\sigma}.
\end{equation}
Here $i$ represents a band index, $\sigma$  represents a spin projection of a fermion, and $\gamma ({\bf k})$  is the Raman form factor,  which keeps track of the polarizations of the incoming and the outgoing light.
The use of light of different polarizations allows the probing of different symmetry channels: $A_{1g}$, $A_{2g}$, $B_{1g}$, and $B_{2g}$.
   Note that the $B_{1g}$ and $B_{2g}$ channels are interchanged when going from the 1-Fe BZ to the 2-Fe BZ,
   because
   the coordinate system is rotated $45^\circ$ to make the $k_x$ and $k_y$ axes coincide with the sides of the square cell.
 Because the wavelength of light used in the experiments is a few orders of magnitude greater than the lattice constant,
the typical values of $v_F p$
are smaller than typical $\Omega$, and it suffices to calculate the susceptibility at
 $p \to 0$,
i.e., compute
 $\chi_R(\Omega)\equiv \chi_R(p \to 0,\Omega)$.

\begin{figure}[htb]
	\centering
		\includegraphics[width=0.45\textwidth]{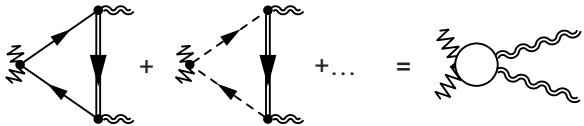}
	\caption{The AL vertex $\Gamma_{tr}$ for the coupling of light to spin fluctuations.
  Single solid and dashed lines represent excitations from one of the two hole bands  and the double solid line represents excitations from one of the two  electron bands. When typical internal frequencies in the triangle made out of fermionic Green's functions are larger than typical frequencies of spin fluctuations,
  $\Gamma_{tr}$ can be approximated by a constant (a circle on the right-hand side of the figure). }
	\label{fig:light}
\end{figure}

 Without the final state interaction, the Raman response involving a pair of spin fluctuations with momenta near $(0,\pi)$ and $(\pi,0)$ (the difference between the centers of electron and hole pockets) is given by the
  diagram shown in Fig. \ref{fig:AL_diag}.  Since light can couple to each hole and electron band, there are several diagrams of this kind with two fermionic lines from one of hole pockets or from one of electron pockets (see Fig.  \ref{fig:triangle}  below).
  The combined contribution from these diagrams  takes the form
\begin{align}\label{eq:AL}
	\chi_R  (\Omega)= &-i\int \frac{\mathrm{d}^2\mathbf{q}\mathrm{d}\nu}{(2\pi)^3} \Gamma^2_{tr,l}(\mathbf{q},\nu) f_l \chi^\mathrm{s}(\mathbf{Q}_l+\mathbf{q},\nu)\nonumber\\
	&\quad \times \chi^\mathrm{s} (\mathbf{Q}_l+\mathbf{q},\nu+\Omega),
\end{align}
where $\Gamma_{tr,l}$ defines the vertex for the coupling between light and spin fluctuations (see Fig. \ref{fig:light}),
 $\chi^\mathrm{s}$ is the propagator of spin fluctuations, $l =1,2$ with $\mathbf{Q}_1 = (\pi,0)$, $\mathbf{Q}_2 = (0,\pi)$, and $f_l$ is the symmetry factor, e.g., $f_l = \sigma^{z}_{ll}$ for $B_{1g}$ geometry.

   The vertex $\Gamma_{tr,l} (\mathbf{q}, \nu)$ is composed out of
  three fermionic Green's functions, the Raman factor $\gamma (\mathbf{k})$, and two vertices for the coupling between fermions and spin-fluctuations:
 $g_{sf} \gamma_{ij} (\mathbf{k}, \mathbf{k}+\mathbf{Q}_l -\mathbf{q})$, where $g_{sf}$ is the spin-fermion coupling (or order of the intra-orbital Hubbard $U$) and $\gamma_{ij} (\mathbf{k}, \mathbf{k}+\mathbf{Q}_l -\mathbf{q})$ are coherence factors
  associated with the transformation from orbital to band basis for fermions  from bands $i$ and $j$ ($i, j =1,2$).
  In a superconducting state
  one must include diagrams containing anomalous Green's functions, which create or destroy particles in pairs, and sum over all allowed combinations of normal and anomalous functions. For any particular combination of normal and anomalous functions, the triangular vertex  takes the form
\bea
\label{eq:Lambda}
&&	\Gamma_{tr,l} (\mathbf{q},\nu)= g^2_{sf} \int \frac{\mathrm{d}^2\mathbf{k}\mathrm{d}\omega}{(2\pi)^3} \gamma_i(\mathbf{k}) \gamma^2_{ij} (\mathbf{k}+\mathbf{Q}_l -\mathbf{q}) \nonumber \\
&& {\tilde G}_i (\mathbf{k},\omega) {\tilde G}_i (\mathbf{k},\omega+\Omega) {\tilde G}_j(\mathbf{k}+\mathbf{Q}_l -\mathbf{q},\omega-\nu),
\eea
where
 ${\tilde G}_{i,j}$ are either normal ($G$) or anomalous ($F$) Green's functions with $i$ from a hole pocket and $j$ from an electron pocket or vice versa.
The  index of the electron band is equal to $l$.

\section{The vertex function in the normal state and the temperature dependence of the $B_{1g}$  susceptibility above $T_c$}
\label{sec:T_dep}

In this section we compute the Raman vertex function in the normal state and obtain the temperature dependence of the real part of the Raman
 susceptibility in the static limit.
Here and in Sec. \ref{sec:Enhancement} we
 focus on $B_{1g}$ scattering geometry with $\gamma (\bf k) \propto \cos{k_x} - \cos{k_y}$  and do not explicitly write symmetry factors in the Raman vertex
  and in the vertices
 relating fermions with spin fluctuations. We will discuss these symmetry factors and different geometries in Sec. \ref{sec:Symmetry}.
     We also neglect for simplicity  the eccentricity of the electron pockets and set all Fermi surfaces to be circles of the same size. This approximation  simplifies calculations, but does not qualitatively affect the temperature dependence compared to a generic case in which the pockets are different.
In the static limit Eq. (\ref{eq:Lambda}) in the normal state reduces to
\begin{align}
	&\Gamma_{tr}(\mathbf{q},\nu_n) \nonumber\\
	&=-A T\sum_{\omega_m}\int \frac{\mathrm{d}^2\mathbf{k}}{(2\pi)^2} \frac{1}{(i\omega_m - \xi^i_\mathbf{k})^2} \frac{1}{i(\omega_m - \nu_n) -
\xi^j_\mathbf{k-q}},
\end{align}
where $A \sim g^2_{sf}$.
 For concreteness we assume that $i$ is a hole band and $j$ is an electron band. The hole and electron dispersions are given by $\xi^i_\mathbf{k}=\mu-\frac{k^2}{2m}=-
        \xi^j_\mathbf{k}$, where $\mu$ is the chemical potential.
         At  $\mathbf{q}=0$ and $\nu_m=0$, $\Gamma_{tr}$ is  given by
\begin{align}
    \Gamma_{tr} &=\frac{A m}{16\pi T}  f\left(\frac{\mu}{2T}\right).
		\nonumber
\end{align}
The scaling function
$f(x)=\tanh(x)/x$
is close to 1
 for $x < 1$, i.e., for $T >\mu/2$. In this temperature range $\Gamma_{tr} \approx A m/(16 \pi T)$ scales as $1/T$. This has been noticed before [\onlinecite{Paul}].
At larger $x$ (smaller $T$), $\Gamma_{tr}$ tends to a constant [\onlinecite{Raman_nematic_khodas}].  At a nonzero $\mathbf{q}$ and $\nu_m$ the expression for $\Gamma_{tr}$ becomes more complex, but as long as $\nu_m = O(T)$ and $|\mathbf{q}| \leq k_F$, the functional form remains the same.

We next compute the convolution of two spin fluctuations in the normal state.   There is no controllable way to obtain the spin-fluctuation propagator starting from the fermion-fermion interaction. The RPA procedure is often used, but it selects  particular series of ladder and bubble diagrams in the particle-hole channel and neglects contributions from the particle-particle channel.  The latter are, however, not small, even at perfect nesting [\onlinecite{chub_review}].  Besides, in a general case of hole and electron pockets of different sizes and geometry, the static propagator of spin fluctuations comes from fermions with energies of order bandwidth, for which the low-energy expansion is not applicable.
 In view of this complication, we adopt the same approach as in earlier works on the spin-fermion model [\onlinecite{acs}] and assume phenomenologically that the static part of the spin-fluctuation propagator has a regular Ornstein-Zernike form $\chi^s_{ij} (\mathbf{q+Q},0) = 1/(q^2 + m^2_s)$, where
 $m_s$ is the inverse magnetic correlation length
 (the overall factor in $\chi^s$ is incorporated into the spin-fermion coupling).
The dynamical part of $\chi^s$, however, comes from low-energy fermions and can be obtained by
  computing the dynamical part of particle-hole polarization bubble made of fermions near a hole and an electron pocket, separated by $\mathbf{Q}$. Then
 \begin{equation}
	\chi^s(\mathbf{Q}+\mathbf{q},\nu_m)= \frac{1}{m^2_s + q^2 + \gamma \Pi_{Q} (\nu_m)},
\label{ch_1}
\end{equation}
where $\gamma = m g_{sf}/(2\pi)$, $g_{sf}$ is the spin-fermion coupling [\onlinecite{acs}],
 and
 $\Pi_Q (\nu_m) = \Pi(\mathbf{Q},\nu_m) - \Pi (\mathbf{Q},0)$, where $\Pi (\mathbf{Q},\nu_m)$ is the dynamical polarization bubble at momentum transfer $\mathbf{Q}$.
  The polarization bubble $\Pi_Q (\nu_m)$ is logarithmic in $\nu_m$ because it is the convolution of fermions from hole and electron bands with opposite sign of the dispersion [\onlinecite{zlatko}]. We computed $\Pi_Q (\nu_m)$ numerically and found that it can be well
   approximated by
  \beq
  \Pi_Q (\nu_m) = \log{\left(3.57 \frac{|\nu_m|}{2\pi T}\right)}
 \label{new_a}
  \eeq
  starting already from the lowest nonzero Matsubara frequency.
  Substituting this into (\ref{ch_1}) and evaluating the convolution of the two dynamical spin susceptibilities with the same momentum and frequency we obtain
\begin{align}
& T\sum_{\nu_m}\int \frac{\mathrm{d}^2 \mathbf{q}}{(2\pi)^2} \chi^s(\mathbf{q+Q},\nu_m) \chi^s(\mathbf{q+Q},\nu_m)\nonumber\\
& \propto T (m_s)^{-2} \left(1 +  \sum_{m \neq 0} \frac{1}{1 + \gamma  (m_s)^{-2}\log{\left(3.57|m|\right)}}\right)
\end{align}
 The coupling constant $\gamma$ cannot be calculated within the theory, but is generally of order 1. Assuming that this is the case, we
  find that the  dominant contribution to the sum over bosonic Matsubara frequencies  comes from the term with $\nu_n =0$, at least when the
   inverse magnetic correlation length $m_s <1$.   The convolution of the two $\chi^s$ then gives, up to a constant prefactor, $T/m^2_s$ [\onlinecite{yama}].
  Combining this with $\Gamma^2_{tr} \propto 1/T$, we obtain that the  contribution to the static $B_{1g}$ Raman susceptibility from the  processes
  involving a pair of spin fluctuations with momenta near $(0,\pi)$ and $(\pi,0)$ and two triple vertices (i.e., from the diagram in
  Fig. \ref{fig:AL_diag})  is given by
  \beq
  \chi_R (\Omega =0) \propto \frac{1}{T m^2_s}
  \label{zz_1}
  \eeq
   Outside the $T$ range near a magnetic transition, the temperature dependence of $m_s$ is weak, and $\chi_R (\Omega =0)$ scales roughly as $1/T$.

\begin{figure}[htb]
	\centering
		\includegraphics[width=0.45\textwidth]{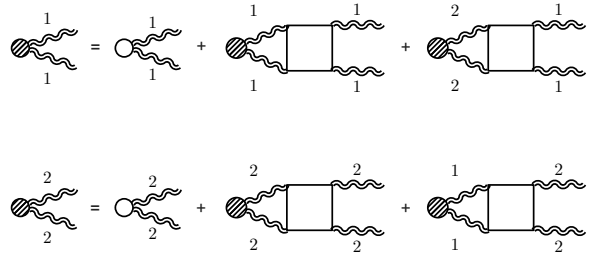}
	\caption{Ladder series of renormalizations of the interaction of light with spin fluctuations with momenta near $\mathbf{Q}_1 =(\pi,0)$ (labeled as $1$) and near $\mathbf{Q}_2 = (0,\pi)$ (labeled as $2$). The interaction vertices are made out of fermions from different bands (see Fig. \protect\ref{fig:int_vertex}).}
	\label{fig:higher_order}
\end{figure}

\begin{figure}[htb]
	\centering
		\includegraphics[width=0.45\textwidth]{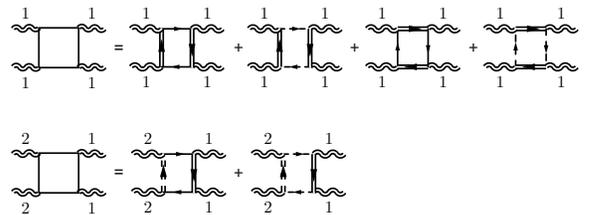}
	\caption{The structure of the vertices for the interaction between spin fluctuations with  near $\mathbf{Q}_1 =(\pi,0)$ (labeled as $1$) and near $\mathbf{Q}_2 = (0,\pi)$ (labeled as $2$).  The single solid and dashed lines represent excitations from one of the two hole bands  and the double solid and dashed lines represent excitations from one of the two  electron bands.}
	\label{fig:int_vertex}
\end{figure}

\subsection{Higher-order contributions to Raman susceptibility}
 We next consider how Eq. (\ref{zz_1})  changes once we include the interactions between pairs of spin fluctuations.
 These interactions either leave bosonic momenta near a particular $\mathbf{Q}_1 = (\pi,0)$ or $\mathbf{Q}_2 = (0,\pi)$, or transfer both momenta from $\mathbf{Q}_1$ to $\mathbf{Q}_2$ or vice versa.
  In the latter case, the process belongs to the umklapp category and is allowed because $2\mathbf{Q}_1 = 2\mathbf{Q}_2$ up to reciprocal lattice vector.  We show the corresponding diagrams in Fig. \ref{fig:higher_order}.    Each interaction vertex
  is given by the convolution of four fermionic propagators (see Fig. \ref{fig:int_vertex}).
    These vertices have been computed in Ref. [\onlinecite{FC}] in the limit when bosonic frequencies
   are set to zero.   We argued above that this approximation is justified for the two-loop diagram, and we now assume that this also holds for higher-order processes.
  We then borrow the results from Ref. [\onlinecite{FC}], which demonstrated that the effective interaction in the $B_{1g}$ channel is a negative (attractive)
   $2g$, which scales with temperature as $1/T^4$.

  We show explicitly how to solve the coupled set of ladder equations for the fully renormalized  AL vertices with momenta $\mathbf{Q}_1$ and $\mathbf{Q}_2$ later in Sec. \ref{sec:Symmetry} (where we compare $A_{1g}$ and $B_{1g}$ channels),  because the set involves both $A_{1g}$ and $B_{1g}$ components. Here  we just present the result: at small $\Omega$,  interactions between spin-fluctuations change the two-loop AL Raman vertex in $B_{1g}$ geometry into
\begin{align}
	\chi_{R} (\Omega) = & \Gamma^2_{tr} \frac{\chi^2 (\Omega)}{1+2g \chi^2 (\Omega)} =  \Gamma^2_{tr} ~\chi_{I-nem} (\Omega),
\label{eq:Raman_RPA_B1g_1}
\end{align}
where  $\chi^2 (\Omega)$ is the short notation for the convolution of two spin-fluctuation propagators with relative frequency $\Omega)$ and we defined
\beq
\chi_{I-nem}  (\Omega) = \frac{\chi^2 (\Omega)}{1+2g \chi^2 (\Omega)}.
\label{zz_3}
\eeq
Using $g \sim 1/T^4$ , $\chi^2 \propto T$, $\Gamma_{tr} \propto 1/T$, and introducing $T_{0}$ as a temperature at which $2 g \chi^2 =1$, we obtain
\begin{align}
\chi_{R}(0, T)\sim &\frac{(m_s)^{-2}}{T-T_{0}} \left(\frac{T^2}{T^2+TT_{0}+T_{0}^2}\right),
\end{align}
 which for $T  > T_{0}$ is rather well approximated  by
 \begin{align}
\chi_R \sim &\frac{1}{T-T_{0}}.
 \end{align}
 In Sec. \ref{sec:Symmetry} we show that singular $1/(T-T_0)$ dependence only holds for the $B_{1g}$ Raman vertex.  In other channels, Raman intensity from the coupling to spin fluctuations either vanishes by symmetry of is substantially  reduced by interaction between spin fluctuations.

 The $1/(T-T_0)$ behavior of the $B_{1g}$ Raman vertex is quite consistent with the experimental observations for NaFe$_{1-x}$Co$_x$As, EuFe$_2$As$_2$,  and Ba(Fe$_{1-x}$Co$_x$)$_2$As$_2$ [\onlinecite{Thorsmolle,Girsh_b,BaFeCoAs1,BaFeCoAs2}].
 In these materials $T_0$ is positive at doping below a certain $x_0$.  This $T_0$ would be the temperature of Ising-nematic instability in the absence of (i) superconductivity and (ii) coupling to phonons.  Superconductivity obviously cuts $1/(T-T_0)$ behavior at $T_c$, if $T_c$ is larger than $T_0$. The coupling to static phonons shifts the
   the coupling $g$ to $g_{eff} = g + (\Gamma_{tr} \lambda_{ph}/C_{ph})^2/2$ [see Eq. (\ref{zz_2})] and hence shifts the  temperature of the Ising-nematic instability to $T_{nem}>T_0$.  We remind that this shift is not present in the "static" $\chi_R $, extracted from the measured Im $\chi_R (\Omega)$ by Kramers-Kronig transformation,
      because the contribution from phonons rapidly drops at nonzero $\Omega$ and is negligibly small for frequencies at which Im $\chi_R (\Omega)$ has been measured.
        As a result, "static" $B_{1g}$  susceptibility extracted from the Raman data increases upon decreasing $T$ but does not diverge even at $T_{nem}$.

 Note also that $T_0$ is positive as long as magnetic order is a stripe, otherwise $g>0$ and the final state interaction reduces rather than enhances the Raman intensity. Recent studies have shown [\onlinecite{cf}] that $g$ does change sign as doping increases, hence one should expect that $T_0$ will change sign from positive to negative above a certain doping. This is fully consistent with the data [\onlinecite{Thorsmolle}].

\section{The resonance in $B_{1g}$ channel below $T_c$}
\label{sec:Enhancement}

\begin{figure}[htb]
	\centering
		\includegraphics[width=0.45\textwidth]{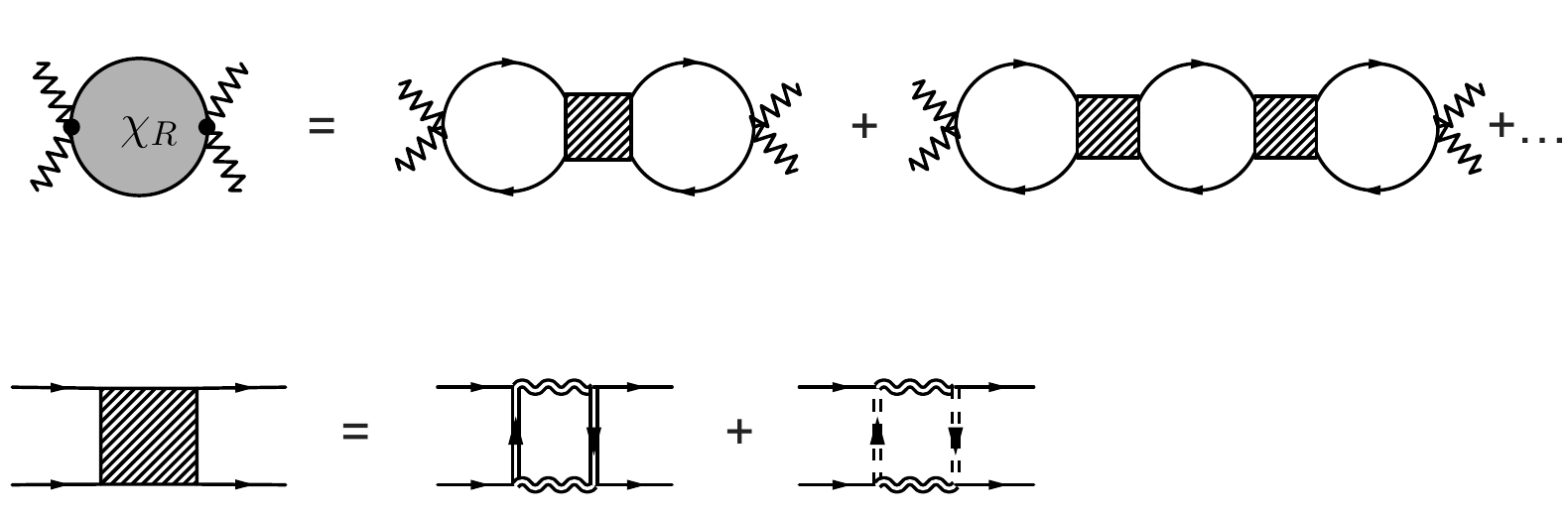}
	\caption{Reinterpretation of  series of AL diagrams for $\chi_R$ with interactions between spin fluctuations  as series consisting of multiple particle-hole bubbles
  with zero momentum and finite frequency transfer (unshaded circles), separated by
effective interactions mediated by spin-fluctuations. Each interaction vertex (shaded rectangle) is the convolution of two fermionic and two bosonic propagators.
  A particular subset of diagrams is shown, with fermions from one of the hole band (solid lines).  Double solid and double dashed lines describe fermions from two electron bands.}
	\label{fig:reinter}
\end{figure}

We now turn to the superconducting state. We first argue in Sec. \ref{sec:Raman_n} that under certain conditions the resonance in $B_{1g}$ Raman response is due to the development of the pole in $\chi_{I-nem} (\Omega)$, given by (\ref{zz_3}).
In Sec. \ref{sec:Raman_3} we consider another scenario for the resonance. Namely, we re-interpret the two-loop diagram as containing two dynamical particle-hole polarization bubbles with zero momentum  transfer, $\Pi_{B_{1g}} (\Omega)$, coupled by an effective interaction mediated by spin fluctuations (see Figs. \ref{fig:reinter_1}, \ref{fig:reinter}).  This effective interaction renormalizes $\lambda$ in Eq. \ref{eq:Raman-RPA} into $\lambda_{eff}$, and we show that $\lambda_{eff}$  becomes negative.  As we said in the Introduction, for negative coupling  the system develops an excitonic resonance in the superconducting state, at a frequency below twice the superconducting gap.

\subsection{Resonance due to the pole in $\chi_{I-nem}$}
\label{sec:Raman_n}

To analyze the form of $\chi_{I-nem}$  at $T=0$ in a superconductor we need to know the form of $\chi^s (\mathbf{Q}+\mathbf{q}, \nu)$ along the real frequency axis. As in earlier works, we assume that the symmetry of the superconducting order parameter is $s^{+-}$.
  Superconductivity does not affect the static form of $\chi^s$ as it generally comes from high-energy fermions, but changes the form of the dynamical term
  $\Pi_{Q} (\nu) = \Pi_{s}(\mathbf{Q},\nu) - \Pi_{s}(\mathbf{Q},0)$ in Eq. (\ref{ch_1})  (converted to real frequencies),
   as this term now contains  the sum of $G_s G_s$ and $F_s F_s$ terms.
		Approximating fermionic dispersion in the same way as before we obtain
\begin{align}\label{eq:Pi0}
	\Pi_{s}(\mathbf{Q},\nu)=\frac{1}{2}\int &\frac{\mathrm{d}^2\mathbf{k}}{(2\pi)^2}\left[\frac{1}{\nu+2 E_\mathbf{k} -i\eta}\right.\nonumber\\
	&-\left.\frac{1}{\nu-2 E_\mathbf{k} +i\eta}\right],
\end{align}
where $E^2_\mathbf{k} = \xi^2_\mathbf{k} + \Delta^2$ and $\xi_{\mathbf{k}} = \mu-\frac{k^2}{2m}$
 (note that we define $\Pi$ without a spin factor of $2$).
In principle, in evaluating $\Pi_s (\mathbf{Q},\nu)$  one has to include also $G_s F_s$ terms and combine renormalizations in the particle-hole and the particle-particle channels because in the superconducting state particles and holes are mixed [\onlinecite{sczhang}].
  Previous work on the subject [\onlinecite{HCW}], however,  has shown that as long as all the interactions are repulsive, the effect of inclusion of these extra terms is minimal in the case of Fe pnictides and merely shifts the resonance frequency (see below) down by a few percentage points.

The straightforward analysis shows that $\operatorname{Im}\Pi_{Q} (\nu)$ vanishes for $|\nu|<2\Delta$ because the excitations are gapped. At $|\nu|=2\Delta$, $\operatorname{Im}\Pi_Q (\nu)$ undergoes a discontinuous jump to a finite value and $\operatorname{Re}\Pi_Q (\mu)$ diverges logarithmically.
 The divergence of the real part of $\Pi_Q (\nu)$ at $\nu =2\Delta$  implies that the denominator  in (\ref{ch_1}) must vanish at some frequency below $2\Delta$, thus creating a pole in $\chi^s(\mathbf{Q}+\mathbf{q},\nu)$.  Specifically, for a given $\mathbf{q}$, $\operatorname{Im} \chi^s(\mathbf{Q}+\mathbf{q},\nu)$ has sharp peak at frequency $\nu_{res}(\mathbf{q})$ and $\operatorname{Re} \chi^s(\mathbf{Q}+\mathbf{q},\nu)$ diverges. This is indeed nothing but the spin resonance in an $s^{+-}$  superconductor [\onlinecite{spin_resonance}].
 Because time-ordered $\Pi_Q (\nu)$ is an even function of $\nu$, it follows that for a given $\mathbf{q}$, time-ordered $\chi^s(\mathbf{Q}+\mathbf{q},\nu)$ has two simple poles at $\nu=\pm \nu_{res}(\mathbf{q})$. Then $\chi^s$ can be written as
\begin{equation}
	\chi^s(\mathbf{Q}+\mathbf{q},\nu)=\frac{a(\mathbf{q},\nu)}{[\nu+\nu_{res}(\mathbf{q})][\nu-\nu_{res}(\mathbf{q})]},
\label{zz_4}
\end{equation}
where $a(\mathbf{q},\nu)$ is some analytic function, which is also even in $\nu$.

We now turn to the  Raman susceptibility from the two-loop diagram, Eq. (\ref{eq:AL}).
 We assume and then verify that the triangular  Raman vertex $\Gamma_{tr}$ can be approximated by a constant and taken out of the integral for $\chi_R$.
 The $1/T$ temperature  dependence of $\Gamma_{tr}$ is obviously cut by $T_c$, i.e., it remains finite at $T=0$.  Whether it can be taken out of the integral over the bosonic frequency is a more subtle issue and we discuss it at the end of this section.

  With $\Gamma_{tr}$ approximated by a constant, the  expression for the Raman susceptibility  takes the  form
\begin{equation}\label{eq:RVV_1}
	\chi_R (\Omega) = \Gamma^2_{tr} \chi^2 (\Omega),
\end{equation}
 where
\begin{equation}\label{eq:RVV}
\chi^2 (\Omega) =  -i \int \frac{\mathrm{d}^2\mathbf{q}\mathrm{d}\nu}{(2\pi)^3} \sigma^z_{ii}  \chi^\mathrm{s} (\mathbf{Q}_i+\mathbf{q},\nu) \chi^\mathrm{s} (\mathbf{Q}_i+\mathbf{q},\nu+\Omega).
\end{equation}
  where we remind that $i=1,2$,  $\mathbf{Q}_1 = (\pi,0), \mathbf{Q}_2 = (0, \pi)$, and $\sigma^z_{ii}$ is present because we consider $B_{1g}$ geometry.

    Substituting $\chi_s$ from (\ref{zz_4}) into Eq. (\ref{eq:RVV}) and evaluating the frequency integral, we obtain, neglecting symmetry factors,
\begin{align}
	\chi^2 (\Omega) = &-\int \frac{\mathrm{d}^2\mathbf{q}}{(2\pi)^2} \frac{a[\mathbf{q},\nu_{res}(\mathbf{q})]}{2\nu_{res}(\mathbf{q}) \Omega}\\
	& \quad\quad \times \[\frac{a[\mathbf{q},\Omega-\nu_{res}(\mathbf{q})]}{\Omega-2\nu_{res}(\mathbf{q})} +\frac{a[\mathbf{q},\Omega+\nu_{res}(\mathbf{q})]}{\Omega+2\nu_{res}(\mathbf{q})}\].
\label{zz_5}
\end{align}

This formula shows that for each momentum $\mathbf{q}$ there is an enhancement of the response at twice the frequency $\nu_{res}(\mathbf{q})$.
We define the minimum value of  $\nu_{res}(\mathbf{q})$ as $\Omega_{mag}$.
   A simple experimentation with the momentum integral shows that Im $\chi^2$ is small at $\Omega<2\Omega_{mag}$, but enhances sharply at $\Omega \geq 2 \Omega_{mag}$. In order to illustrate this effect more concretely, we adopt a simple model for the dispersion of the pole. Namely, we set
 $\nu_{res}(\mathbf{q}) =\Omega_{mag} +\alpha q^2$. Integrating in
 Eq. (\ref{zz_5}) over $q$ and substituting the result into (\ref{eq:RVV_1}) we obtain
\begin{equation}
	\chi_R (\Omega)= \frac{\Gamma^2_{tr} a^2}{8 \pi \alpha \Omega^2} \log{\frac{4\Omega^2_{mag}}{4\Omega^2_{mag}-\Omega^2}} .
\end{equation}

  We see that in the two-loop approximation,
    $\operatorname{Im}\chi_{R}(\Omega)$ undergoes a jump from zero to a finite value at $\Omega = 2 \Omega_{mag}$. The real part of the Raman susceptibility $\operatorname{Re}\chi_{R}(\Omega)$ diverges logarithmically at this frequency.   Below we  verify this result by evaluating (\ref{eq:RVV}) numerically.

We next follow the same path as in the normal state and  include higher-order diagrams (Fig. \ref{fig:Box}) with the interactions between the two spin fluctuations, i.e.,
 replace the two-loop result $\chi_R = \Gamma^2_{tr} \chi^2 (\Omega)$ by
 \begin{equation}
 \chi_R  (\Omega) = \Gamma^2_{tr} \chi_{I-nem} (\Omega) = \Gamma^2_{tr} \frac{\chi^2 (\Omega)}{1+2g \chi^2 (\Omega)}.
 \label{zz_6}
 \end{equation}
    Because Re $\chi^2$ diverges upon approaching $2 \Omega_{mag}$ from below, Im $\chi^2$ vanishes below $2\Omega_{mag}$, and $g <0$,
  the full $\mathrm{Im} \chi_R (\Omega)$  has a true pole at some frequency $ \Omega = \Omega_{res,1} < 2 \Omega_{mag}$.

We now verify the approximation  that $\Gamma_{tr}$ can be taken out of the integral over the bosonic frequency $\nu$.
  The triangular vertex contains one internal frequency $\omega$ and two external ones:  $\Omega$,  at which we probe the Raman response, and the bosonic frequency $\mu$. For $\Omega$ we take the resonance frequency $\Omega_{res,1} < 2 \Omega_{mag}$.  Typical bosonic frequency $\nu \sim \Omega_{mag}$ and typical $\omega$ is $\Delta$.
  Obviously then, the AL vertex $\Gamma_{tr}$ is independent on $\nu$ if $\Omega_{mag}$ is much smaller than $\Delta$, i.e., when internal energy in the AL  diagram  made out of three fermionic Green's functions is much larger than both external frequencies.

  The condition $\Omega_{mag} \ll \Delta$ is satisfied when the inverse magnetic correlation length $m_s$ is small enough because $\Omega_{mag} \propto m_s$ [\onlinecite{ac_1}].  As we said in the Introduction, in  Fe-based materials, where $B_{1g}$ resonance has been observed, the situation is somewhat different: neutron scattering data for NaFe$_{1-x}$Co$_x$As with $x=0.045$ show [\onlinecite{neutron_0.045}] that $\Omega_{mag}\approx 7$ meV, while $\Delta = 5-5.5$ meV [\onlinecite{prx}], i.e. $\Delta$ is somewhat smaller. In this situation, there is no good reason to treat $\Gamma_{tr}$ as a constant.

In the next section we  analyze another scenario, which is justified in the opposite limit when
$\Omega_{mag}$ is much larger than $\Delta$.

\subsection{Another interpretation of AL contribution to the $B_{1g}$ Raman response  below $T_c$.}
\label{sec:Raman_3}

In this section we look at the second-order AL diagram from Fig. \ref{fig:AL_diag} through different lenses.  Namely, we abandon the approximation in which $\Gamma_{tr}$ is treated as a constant and instead use the fact that on both ends incoming and outgoing fermionic momenta are identical (either ${\mathbf k}$ or ${\mathbf p}$), while frequencies differ by $\Omega$, and re-interpret this diagram as consisting of the product of two particle-hole polarization operators at zero transferred momentum and finite frequency, $\Pi_{B_{1g}}(\Omega)$ [the same as in Eq.  (\ref{eq:Raman-RPA})], separated by magnetically mediated effective interaction $\lambda_{mag}$ (see Fig. \ref{fig:reinter}). The latter is the convolution of two fermionic and two bosonic propagators.
 Viewed this way, the two-loop AL diagram has the same structure as the two-loop diagram from RPA series in Eq.  (\ref{eq:Raman-RPA}).
  Accordingly, $\lambda_{mag}$ and the bare $\lambda$ are combined into $\lambda_{eff} = \lambda + \lambda_{mag}$.  If the combined $\lambda_{eff}$ is negative,  $1 + \lambda_{eff} \Pi_{B_{1g}}(\Omega)$ necessary vanishes at some frequency below $2\Delta$ because Im $\Pi_{B_{1g}}(\Omega)$  vanishes below $2\Delta$ and Re $\Pi_{B_{1g}}(\Omega)$ diverges as $1/\sqrt{4\Delta^2 - \Omega^2}$ when $|\Omega|$ approaches $2\Delta$ from below. The vanishing of $1 + \lambda_{eff} \Pi_{B_{1g}}(\Omega)$ implies that Raman intensity has an excitonic  resonance at $\Omega = \Omega_{res,2}$.

 The representation of the two-loop AL diagram from Fig. \ref{fig:AL_diag} as $\lambda_{mag} \Pi^2 _{B_{1g}}(\Omega)$ with a constant $\lambda_{mag}$
 is again an approximation because the result for the convolution of two fermionic and two bosonic propagators generally depends on external momentum in frequency. The singular behavior of the particle-hole polarization bubble $\Pi_{B_{1g}}(\Omega)$ at $\Omega \approx 2\Delta$ comes from internal frequencies near $\pm \Delta$.  Internal frequencies in the fermionic-bosonic loop for $\lambda_{mag}$ are of order $\Omega_{mag}$.  If $\Omega_{mag}$ is much larger than $\Delta$,  a typical internal frequency is much larger than a typical external frequency. The latter can then be sent to zero, in which case $\lambda_{mag}$ becomes a constant.
  The frequency  $\Omega_{mag} <2\Delta$ and hence it can be at most twice $\Delta$. But number-wise this may be sufficient to treat $\lambda_{mag}$ as a constant.
    The same distinction holds for internal/external momenta, and the result is that, to the same accuracy,
  $\lambda_{eff}$ can be evaluated by placing external momenta on the Fermi surface.

We compute $\lambda_{eff}$ first in the normal state and then in a superconductor, assuming formally that $\Omega_{mag} \gg \Delta$.
To simplify calculations, we set $\mu=0$, i.e., assume that the  size of hole/electron pockets is  infinitesimally small.
  The argument is that, if $\lambda_{eff}$ has a definite sign in this limit, then, by continuity, the sign should remain the same at a small but finite  $\mu$.
  %

In the normal state,  the coupling $\lambda_{mag}$ is given by
\begin{equation}
\lambda_{mag} =- g^2_{sf} \int \frac{\mathrm{d}^2\mathbf{q}\mathrm{d}\nu}{(2\pi)^3}G^2(\mathbf{q},\nu) \left(\chi^s(\mathbf{q},\nu)\right)^2,
\label{zz_7}
\end{equation}
where $\nu$ is the Matsubara frequency.
For definiteness, we take fermions from one of the electron bands, i.e., use $G(\mathbf{q},\nu) = 1/(i\nu - q^2/(2m))$. We verified that $\lambda_{mag}$ does not change if we instead take fermions from the hole band. For the dynamical spin susceptibility we use Landau-overdamped form extended to Matsubara frequencies: $\chi(\mathrm{q},\nu)=1/(q^2+m_s^2+|\nu|/\nu_0)$, where
 $\nu_0$ is a positive constant.

\begin{figure}[htb]
	\centering
		\includegraphics[width=0.45\textwidth]{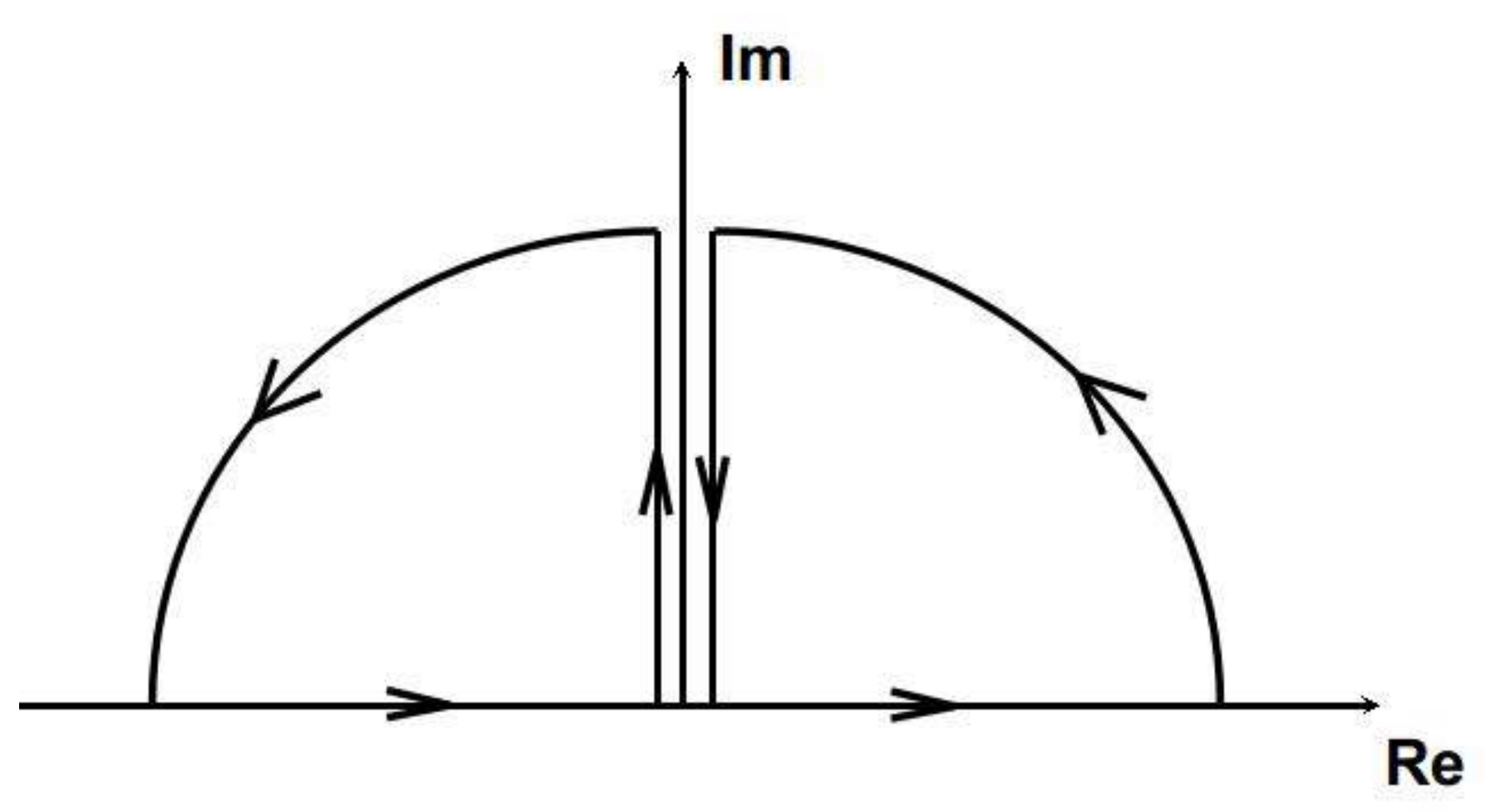}
	\caption{Contour of integration over frequency $\nu$ in Eq. \protect\ref{zz_9}.}
	\label{fig:contour}
\end{figure}

 Substituting the forms of bosonic and fermionic propagators into (\ref{zz_7}) we obtain
 \begin{align}
&\lambda_{mag} =-g^2_{sf} \int \frac{\mathrm{d}^2\mathbf{q}\mathrm{d}\nu}{(2\pi)^3} \frac{1}{[i\nu-\mathbf{q}^2/(2m)]^2}\frac{1}{(\mathbf{q}^2+m_s^2+|\nu|/\nu_0)^2}\nonumber\\
&= \frac{\nu_0^2 m}{(2\pi)^2}\int_0^\infty\mathrm{d}x\int_{-\infty}^{\infty}\mathrm{d}\nu \frac{1}{(\nu +i x)^2}\frac{1}{(|\nu|+\nu_0 m_s^2+\nu_0 m x)^2}.
\label{zz_9}
\end{align}
 The double pole in the fermionic Green's function is located at $\nu = -i x$. It is then convenient to evaluate the frequency integral by closing the integration contour over the upper half plane of complex $\nu$ (see Fig. \ref{fig:contour}).
   The integrand vanishes at $|\nu| \to \infty$, and if $\chi^s$ was an analytic function of $\nu$, $\lambda_{mag}$ would be zero. But this is not the case because the Landau damping term contains $|\nu| = \sqrt{\nu^2}$, which is non-analytic function of $\nu$ along imaginary axis in both half planes. Choosing the integration contour as shown in Fig. \ref{fig:contour} and using $|i z + \epsilon| = iz \sgn{\epsilon}$, we obtain after a simple algebra
\bea
\lambda_{mag}&=& - i g^2_{sf}~ \frac{\nu_0^2 m}{(2\pi)^2}\int_0^{\infty} \mathrm{d}x\int_0^\infty \mathrm{d}z  \frac{1}{(z+ x)^2} \nonumber \\
&& \times\left[\frac{1}{(i z+\nu_0 m_s^2+\nu_0 m x)^2} - \frac{1}{(-i z+\nu_0 m_s^2+\nu_0 m x)^2}\right] \nonumber \\
&&=-\frac{4 g^2_{sf}\nu_0^2 m}{(2\pi)^2} \nonumber \\
&& \times\int_0^{\infty} \mathrm{d}x\int_0^\infty\mathrm{d}z \frac{z (\nu_0 m_s^2+\nu_0 m x) }{(z+ x)^2 [z^2+\nu^2_0 (m_s^2+ m x)^2]^2}.
\eea
 The integrand is positive, hence $\lambda_{mag} <0$. Estimating the integral, we obtain $\lambda_{mag} \propto 1/m^4_s$, i.e., $\lambda_{mag}$ strongly increases near the magnetic instability.

In the superconducting state, we represent spin-fluctuation propagator by Eq. (\ref{zz_4}), i.e., by $\chi(\mathrm{q},\nu) = a/[\nu^2 + \nu^2_{res} (\mathbf{q})]$,
and use $\nu_{res} (\mathbf{q}) = \Omega_{mag} + \alpha q^2$.  We assume and then verify that typical $\nu$ in the integral for $\lambda_{mag}$ are of order $\Omega_{mag}$.
 Because we assume $\Omega_{mag} \gg \Delta$, we can still use normal state Green's functions for fermions   Substituting into (\ref{zz_7}) we obtain
 \begin{align}
&\lambda_{mag} = \nonumber \\
&-g^2_{sf} \int \frac{\mathrm{d}^2\mathbf{q}\mathrm{d}\nu}{(2\pi)^3} \frac{1}{[i\nu-\mathbf{q}^2/(2m)]^2}\frac{a}{[\nu^2 + (\Omega_{mag} + \alpha q^2)^2]^2}\nonumber\\
&= \frac{g^2_{sf}}{4\pi^2} \frac{m a}{\Omega^4_{mag}}\int_0^\infty\mathrm{d}x\int_{-\infty}^{\infty}\mathrm{d}\nu \frac{1}{(\nu +i x)^2}\frac{1}{[\nu^2+(1 + \beta x)^2]^2}.
\label{zz_9_1}
\end{align}
 where $\beta = 2m \alpha$ is a dimensionless parameter.   The integrand, viewed as a function of $\nu$,  contains two double poles  in the lower-half plane, at $\nu = -ix$ and at $\nu = -i (1 + \beta x)$, and the double pole in the upper half-plane, at $\nu = i (1 + \beta x)$. The last two double poles come from $\chi^2$. Evaluating the frequency integral by standard means, we
  obtain after simple algebra that 
  \begin{align}
\lambda_{mag} = -\frac{g^2_{sf}}{8\pi} \frac{m a}{\Omega^4_{mag}}\int_0^\infty\mathrm{d}x \frac{3 + x(1+3 \beta)}{(1 +\beta x)^3 [1 + (1 + \beta) x]^3}.
\label{zz_9_2}
\end{align}
The integrand is positive for all  $x>0$, hence $\lambda_{mag} <0$, like in the normal state. Furthermore,  because $\Omega_{mag} \propto m_s$, we still have $\lambda_{mag} \propto 1/m^4_s$.

\begin{figure}[htb]
	\centering
		\includegraphics[width=0.45\textwidth]{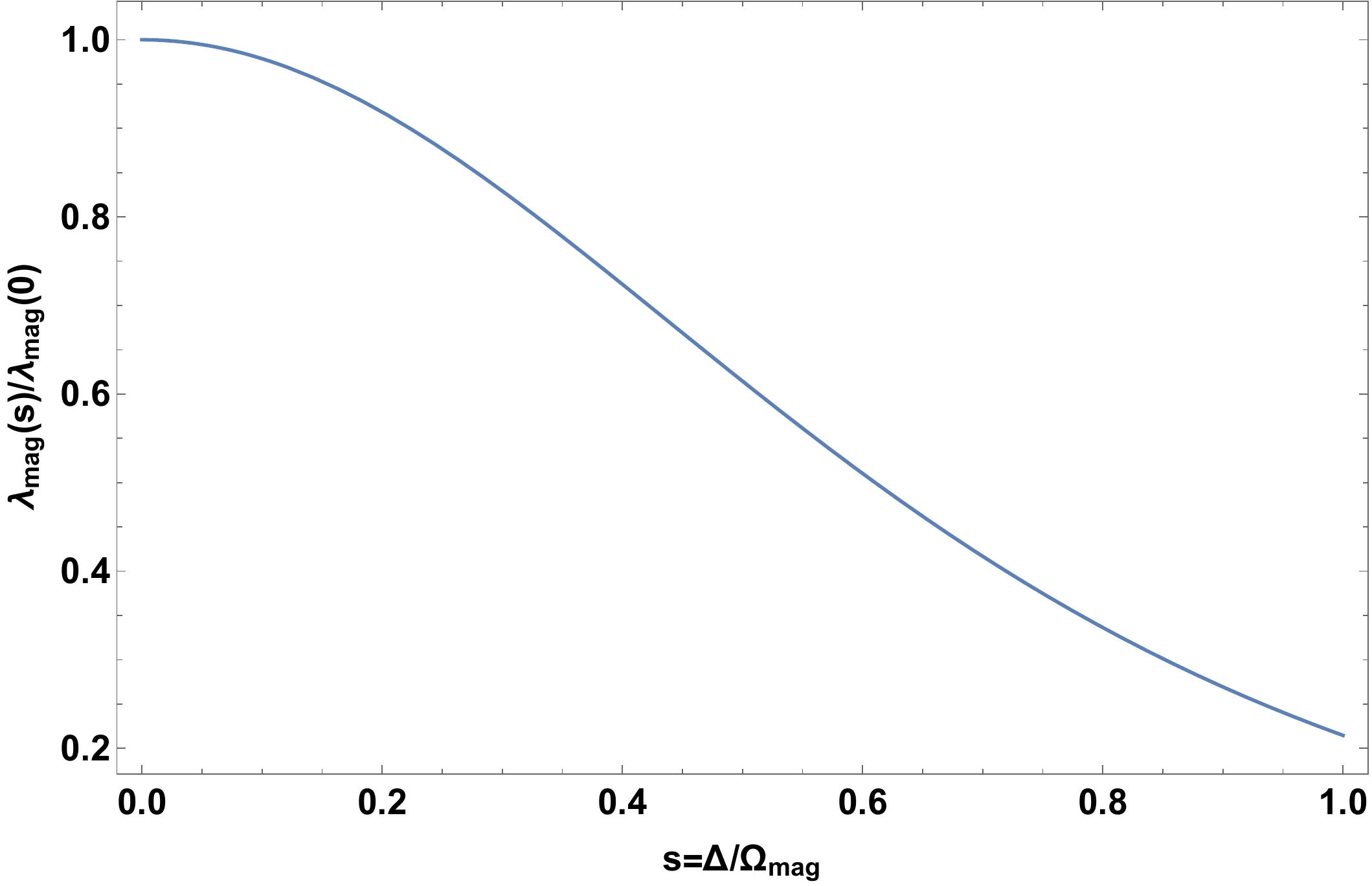}
	\caption{The dependence of the effective spin-mediated four-fermion interaction $\lambda_{mag} (s)$ on $s=\Delta/\Omega_{mag}$, see Eq. (\protect\ref{zz_9_3}).
  We set $\beta =1$ for definiteness.}
	\label{fig:ac}
\end{figure}

Like we said, the condition $\Omega_{mag} \gg \Delta$ is not realized because $\Omega_{mag} \leq 2\Delta$.
 To estimate how $\lambda_{mag}$ changes when $\Omega_{mag}$ and $\Delta$ become comparable, we use the fact that $2\Delta$ singularity in the particle-hole bubble comes from fermions with frequencies near $\pm \Delta$ and evaluate $\lambda_{mag}$ in the superconducting state for the case when the frequencies of the two bosonic propagators differ by $\Omega_m = 2\Delta$. The calculation is straightforward, and the result is that $\lambda_{mag}$ becomes a function of $s = \Delta/\Omega_{mag}$.  The dependence on $s$ is given by
  \begin{align}
&\lambda_{mag} (s) = -\frac{g^2_{sf}}{8\pi} \frac{m a}{\Omega^4_{mag}}  \nonumber \\
&\times \int_0^\infty\mathrm{d}x \frac{[(1 + (1 +\beta)x] [3 + x(1+3 \beta)]-s^2}{(1 +\beta x) [(1 + \beta x)^2 + s^2]  \{[1 + (1 + \beta) x]^2 +s^2\}^2}
\label{zz_9_3}
\end{align}
 We plot $\lambda_{mag} (s)$ in Fig. \ref{fig:ac}. We see that $\lambda_{mag} (s)$ drops when $s = \Delta/\Omega_{mag}$ increases, but the ratio $\lambda_{mag}(s)/\lambda_{mag} (0)$ remains of order one when $\Delta$ and $\Omega_{mag}$ become comparable.

Combining this last observation with the fact that $\lambda_{mag} (0) \propto 1/m^4_s$, we conclude that for small enough $m_s$,
 $|\lambda_{mag}|$   is definitely larger than the bare interaction $\lambda$, hence $\lambda_{eff} = \lambda + \lambda_{mag}$ is negative, no matter what is the sign of $\lambda$.

One can go a step further and  add to the interaction vertex in  Fig. \ref{fig:reinter} (the shaded rectangle) the renormalizations coming from fermions with energies higher than $\Omega_{mag}$.  These terms renormalize the convolution of two spin-fluctuation propagators $\chi^2 (\Omega)$ into $\chi_{I-nem} = \chi^2/[1+2g \chi^2 (\Omega)]$
 and hence add the same denominator to $\lambda_{mag}$. As the consequence, $\lambda_{mag}$ diverges already at the Ising-nematic instability, before $m_s$ vanishes.

 By continuity, we assume that $\lambda_{mag}$ remains negative also for  finite  hole and electron pockets.  Using further RPA form for
  the Raman intensity, Eq.  (\ref{eq:Raman-RPA}),  we obtain that $\chi_R (\Omega)$ has the singularity at $\Omega = \Omega_{res,2} < 2\Delta$, at which
   $1 + \lambda_{eff} \Pi_{B_{1g}}(\Omega) =0$.

 Finally, we briefly comment on the difference between our analysis and earlier phenomenological consideration of the bi-linear coupling between $B_{1g}$ orbital order parameter $\Delta_{oo} = \sum_k <c^\dagger_k c_k \cos 2 \theta_k>$  and  $\Delta^2_1 -\Delta^2_2$,
 where $\Delta_1$ and $\Delta_2$ are spin-fluctuation fields
  with momenta near $\mathbf{Q}_1$ and $\mathbf{Q}_2$ [\onlinecite{BaFeCoAs2,Gallais_1}].
 In the microscopic calculation [\onlinecite{FC}]  such term appears in the Landau free energy once we
 introduce $\Delta_{1,2}$ and $\Delta_{oo}$ as order parameter fields, bi-linear in fermions, and perform Hubbard-Stratonovich transformation from fermions to bosonic  collective variables.  The prefactor $A$ for $\Delta_{oo} (\Delta^2_1-\Delta^2_2)$ term in the Landau functional is given by the same triangular diagram as AL vertex, and has a finite value (i.e., $A \sim \Gamma_{tr}$).  At the first glance,  we can identify $\Delta^2_1-\Delta^2_2$ with the propagator of an Ising-nematic field and obtain the correction to the prefactor for  $\Delta^2_{oo}$  in the form $- A^2 \chi_{I-nem}$. Because the bare prefactor is $\Pi^{-1}_{B_{1g}} + \lambda$,  $\lambda_{eff} = \lambda - A^2 \chi_{I-nem}$. At the second glance, however, we note that the Landau functional in  terms of $\Delta_1$ and $\Delta_2$ is not the same as Landau functional expressed in terms of the Ising-nematic field.  To obtain the latter one has to do a second Hubbard-Stratonovich transformation to the composite Ising-nematic bosonic field $\Delta_{I-nem}$ and integrate over the primary fields $\Delta_1$ and $\Delta_2$.  Only then one can extract the bi-linear coupling between orbital and Ising-nematic order parameters.   Another way to see that $A^2 \chi_{I-nem}$ with $A \sim \Gamma_{tr}$ is not the correction to $\lambda$ is to notice that this expression is the full result for the Raman bubble rather than for the effective interaction between fermions from the two particle-hole bubbles.

\subsection{Comparative analysis of the two scenarios}
\label{sec:comp}

Combining the results of the last two sections, we see that the resonance in $\chi_R (\Omega)$ holds independent of whether $\Omega_{mag}$ is larger or smaller than $\Delta$, but the physics is different in the two cases. When $\Omega_{mag}$ is smaller than $\Delta$,  the resonance has purely magnetic origin and comes from the pole in $\chi_{I-nem}$ at $\Omega = \Omega_{res,1} \leq 2 \Omega_{mag}$.  For this resonance, the role of fermions is to provide some regular coupling, $\Gamma_{tr}$, between incoming and outgoing light and a pair of spin fluctuations with momenta near $\mathbf{Q}_1$ or $\mathbf{Q}_2$. When $\Omega_{mag} > \Delta$, the resonance  comes from fermions and is due to singular behavior of particle-hole polarization bubble $\Pi_{B_{1g}} (\Omega)$ at $\Omega =2\Delta$.  The resonance occurs at a frequency $\Omega = \Omega_{res,2} \leq 2 \Delta$.  Spin fluctuations are again crucial, but now their role is to provide strong attractive interaction between fermions which make particle-hole bubbles.

We treated the two singularities in $\chi_R (\Omega)$  independent of each other chiefly to demonstrate that they come from two different pieces of physics.  Such a treatment, however, is justified only if $\Omega_{res,1}$ and $\Omega_{res,2}$  are well separated.
  In our case, $\Omega_{res,1} \leq 2\Omega_{mag} < 4 \Delta$, while
$\Omega_{res,2} \leq 2 \Delta$.  How well $\Omega_{res,1}$ and $\Omega_{res,2}$ are separated then depends on the strength of various interactions and on the value of magnetic correlation length.  Like we said, in FeSCs that we analyze, $\Omega_{mag}$ and $\Delta$ are not far from each other. In this case the resonance likely has a dual origin.   In NaFe$_{1-x}$Co$_x$As with $x=0.045$,  the $B_{1g}$ peak is seen at 7.1 meV, which is below both $2\Delta$ and $2\Omega_{mag}$.  This is consistent with  the  dual origin of the resonance.

\subsection{Numerical evaluation of the AL diagrams}
\label{sec:Raman_2}

In this section we present the results of numerical evaluation of AL contributions to $B_{1g}$ Raman intensity below $T_c$  first in the two-loop approximation and then including the interaction between spin fluctuations.  We first compute the AL vertex assuming that $\Gamma_{tr}$ can be approximated by a constant and then present the results of the explicit calculation of the two-loop AL diagram for $\chi_R (\Omega)$.

The first step for numerical evaluation of $\chi_R$ is to calculate the time-ordered polarization function $\Pi_{s,ij}$
with fermions lines from bands $i$ and $j$.
The bare spin response can be obtained as a time-ordered average of spin operators over a non-interacting ground state. In the FeSCs, the response is largest near the nesting momenta $\mathbf{Q}_1=(\pi,0)$ or $\mathbf{Q}_2=(0,\pi)$, which connect one hole pocket and one electron pocket. Since we are solely interested in evaluating the function near those momenta we will only consider band combinations of one hole and one electron pocket and drop the band indices from here on. We evaluate the function at momentum $\mathbf{Q}+\mathbf{q}$, where $\mathbf{Q}$ is either $\mathbf{Q}_1$ or $\mathbf{Q}_2$, whichever is appropriate.

For concreteness, we assume parabolic dispersions for the hole and electron pockets of the form $\xi_\mathbf{k}=\mu-\frac{k^2}{2m_h}$ and $\xi_{\mathbf{k}+\mathbf{Q}}=-\mu+\frac{k_x^2}{2m_x}+\frac{k_y^2}{2m_y}$, respectively. We evaluate all quantities in units of the gap $\Delta$ and for numerical parameters we choose $\mu=2\Delta$, $m_h\approx 0.056 \Delta^{-1}$ ($k_F=0.15 \pi/a$), $m_x=m_h/1.27$, and $m_y=m_x/0.3787$. These values approximately fit the bands and Fermi surfaces reported in ARPES measurements [\onlinecite{ARPES_NaFeCo0.05As}]
of NaFe$_{1-x}$Co$_x$As
for  $x=0.05$ (of the two hole bands, we fitted the one with the largest Fermi surface).
For numerical convergence we included a finite broadening $\eta=\Delta/100$.

\begin{figure}[htb]
	\centering
		\includegraphics[width=0.45\textwidth]{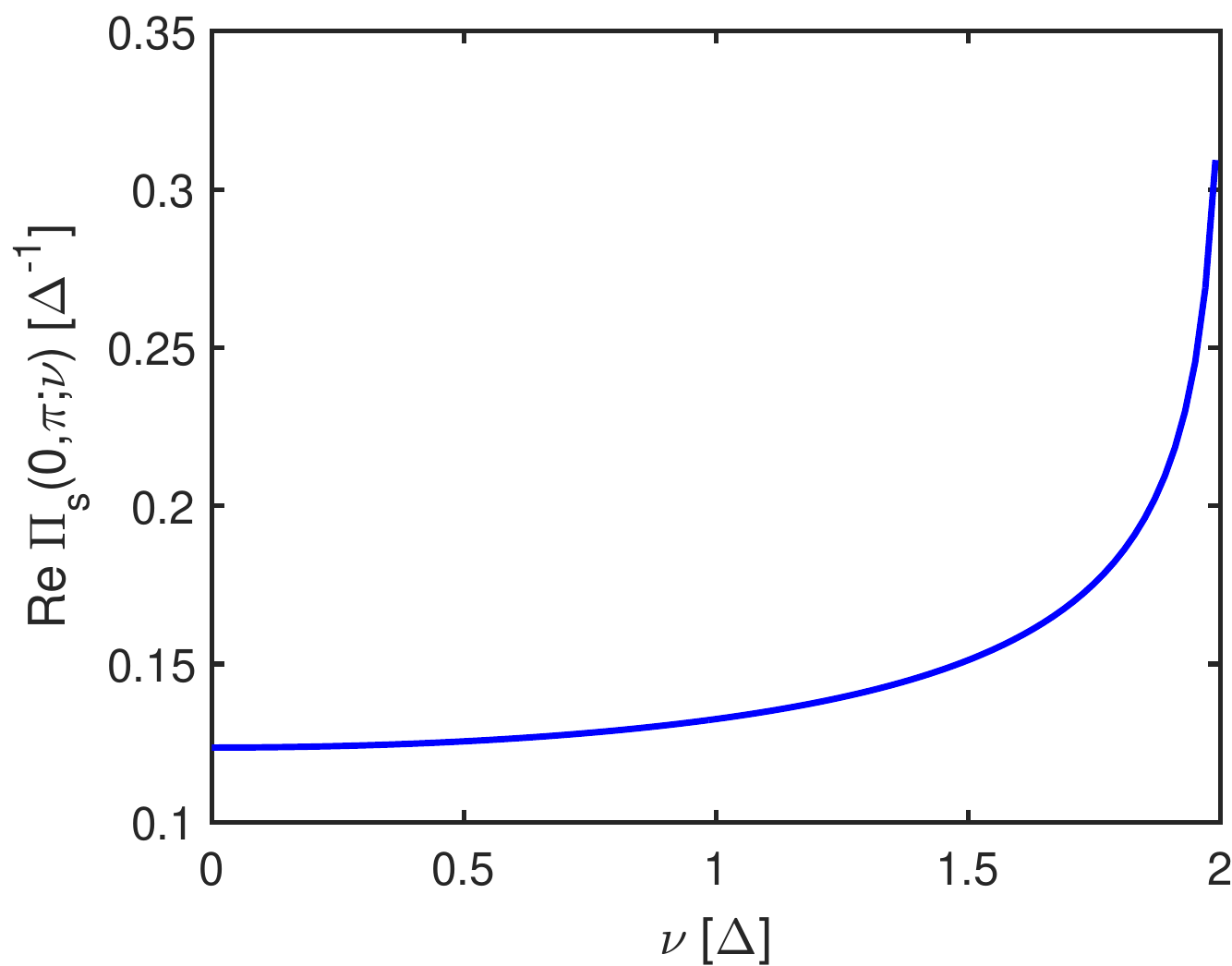}
	\caption{Frequency dependence of the bare spin polarization operator with momentum $(0,\pi)$.}
	\label{fig:Pi_0_w}
\end{figure}

\begin{figure}[htb]
	\centering
		\includegraphics[width=0.45\textwidth]{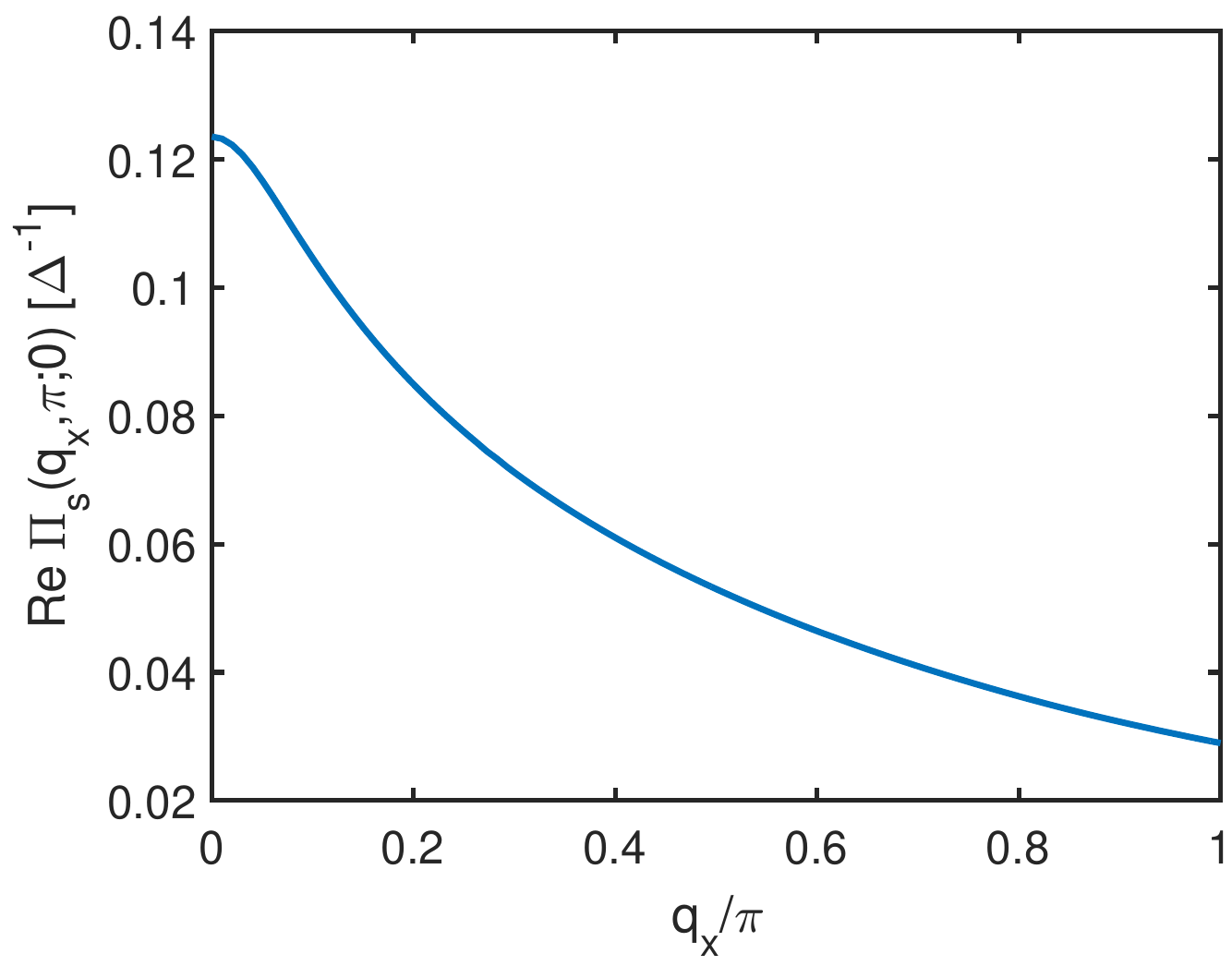}
	\caption{Momentum dependence of the static spin polarization operator with momentum near $(0,\pi)$. We have chosen momenta to be ${\bf q} + (0,\pi)$.
The dependence on $q_x$ is shown.  The dependence on the polar angle of ${\mathbf q}$ is nonzero but rather weak.}
	\label{fig:Pi_0_static}
\end{figure}

The general behavior of $\Pi_s$ can be seen in Figs. \ref{fig:Pi_0_w} and \ref{fig:Pi_0_static}. The first one shows a frequency sweep of the real part at $\mathbf{q}=0$ and the divergence at $2\Delta$ is clearly seen. The imaginary part (not shown) vanishes as $\eta \rightarrow 0$. This behavior holds unless $\mathbf{q}$ is so large that the normal-state FSs no longer intersect due to the shift. The second plot shows an example of the $\mathbf{q}$ dependence at $\nu=0$. Although the function is anisotropic due to the eccentricity of the electron Fermi surface, the qualitative behavior is the same regardless of the polar angle. It is particularly important to emphasize that the function decreases monotonically with increasing $|\mathbf{q}|$.

In the numerical analysis, it is easier to deal with the effective interaction in the spin channel
 $U_{eff}$ rather than the spin susceptibility.  In the RPA
\beq
   U_{eff}(\mathbf{Q}+\mathbf{q},\nu) = \frac{u}{2} \frac{1}{1 - u \Pi_s (\mathbf{Q}+\mathbf{q},\nu)},
   \label{new_3}
 \eeq
 where $u>0$ is the bare fermion-fermion interaction ($= U$ in the Hubbard model). The effective interaction is related to the spin susceptibility by
 $ U_{eff}(\mathbf{Q}+\mathbf{q},\nu)=[u+u^2 \chi^s(\mathbf{Q}+\mathbf{q},\nu)]/2$,
  so the two functions have the same pole structure and differ only by a constant shift $u$, which near a magnetic instability is small compared to the second term. The $u^2/2$ factor in front of $\chi^s$ is the same factor as in (\ref{ch_1}).

\begin{figure}[htb]
	\centering
		\includegraphics[width=0.45\textwidth]{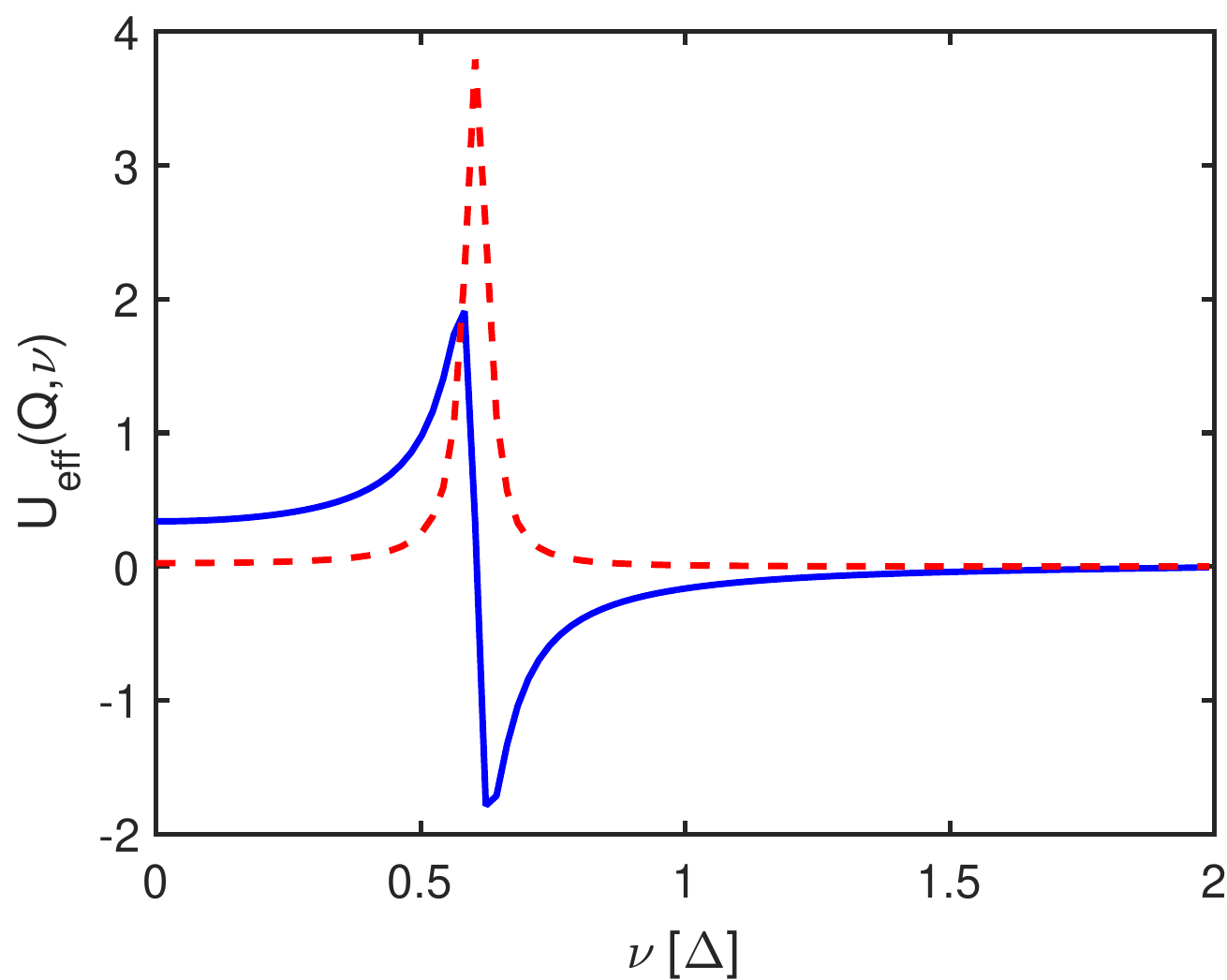}
	\caption{Frequency dependence of the real (solid blue line) and imaginary (dashed red line) parts of the effective interaction  (in arbitrary units).}
	\label{fig:Vs}
\end{figure}

Figure \ref{fig:Vs} shows $U_{eff}$ at $\mathbf{q}=0$ as a function of $\nu$. The real part diverges at the resonance frequency while the imaginary part has a sharp peak which in the limit of $\eta\rightarrow 0$ becomes a $\delta$ function. For the numerical calculations we have set $u \approx 7.9 \Delta$, which determines $\Omega_{mag}\approx 0.6 \Delta$.

Now we are ready to evaluate the Raman response (\ref{eq:RVV}). By using the spectral representation, the imaginary part of the response function can be equivalently calculated as
\begin{align}\label{eq:Im_Raman}
	\operatorname{Im}\chi_{R} (\Omega)& \propto \int \frac{\mathrm{d}^2\mathbf{q}}{(2\pi)^2}\int_0^\Omega\frac{\mathrm{d}\nu}{\pi} \operatorname{Im} U_{eff}(\mathbf{Q}+\mathbf{q}, \nu)\nonumber\\
	&\quad\times\operatorname{Im} U_{eff}(\mathbf{Q}+\mathbf{q}, \nu-\Omega).
\end{align}
The advantage of this form is that it only requires knowledge of the function in a finite range of $\nu$. The real part can then be calculated by using the $KK$ transformation.

\begin{figure}[htb]
	\centering
		\includegraphics[width=0.45\textwidth]{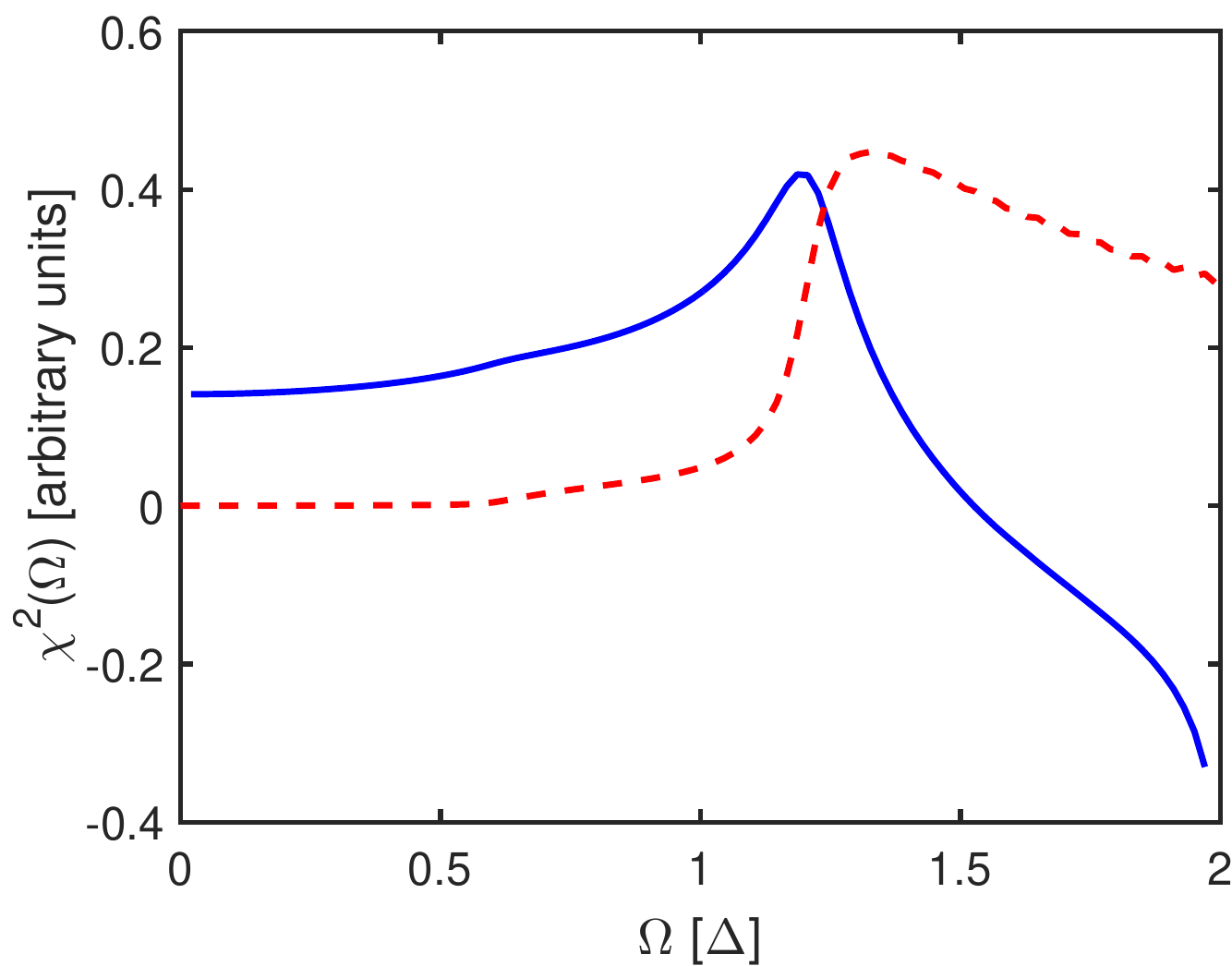}
	\caption{Contribution to Raman response from spin fluctuations. The solid blue and dashed red lines indicate the real and imaginary parts, respectively.}
	\label{fig:Raman}
\end{figure}

 Because $\operatorname{Im} U_{eff}(\mathbf{Q}+\mathbf{q},\nu)$ is peaked at the resonant frequencies corresponding to each momentum $\mathbf{q}$, $\operatorname{Im}\chi_{R} (\Omega)$ can be seen as a convolution of many of these peaks. The result of the computation is shown in Fig. \ref{fig:Raman}. We see that the imaginary part starts small and undergoes a jump at $\nu\approx 1.2\Delta=2\Omega_{mag}$, (compare to Fig. \ref{fig:Vs}). Its value at higher frequencies comes from contributions from $\mathbf{q}\neq 0$, corresponding to excitons of higher energies. This jump is not sharp in the numerical calculations because of the finite value of $\eta$ in (\ref{eq:Pi0}).

\begin{figure}[htb]
	\centering
		\includegraphics[width=0.45\textwidth]{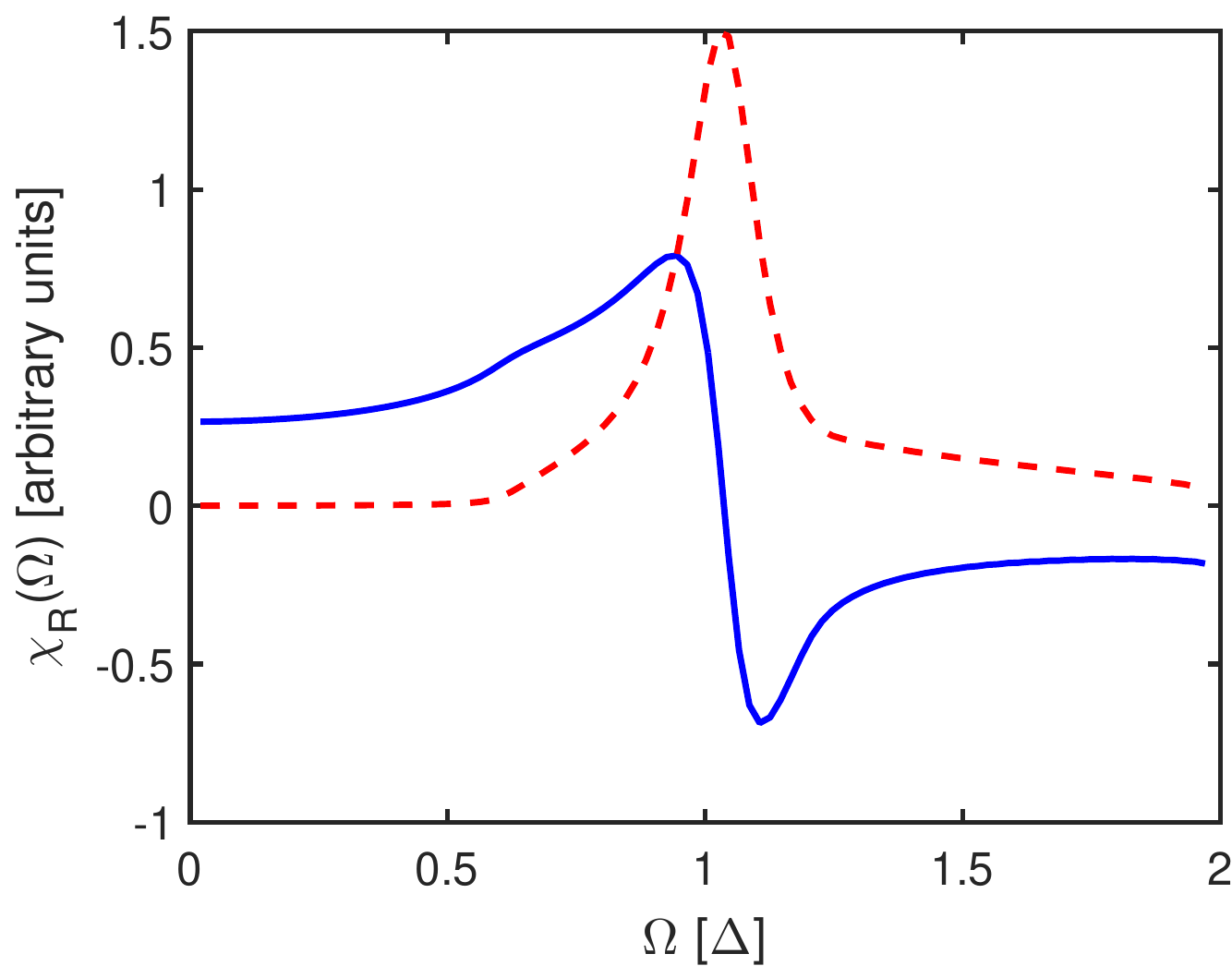}
	\caption{The full Raman susceptibility, which includes interactions between spin fluctuations. The solid blue line indicates the real part and the dashed red line the imaginary part of the function. We set
	$g=1.66$.}
	\label{fig:Raman_full}
\end{figure}

We next consider the effect of the finite state interaction, i.e., include higher order diagrams (Fig. \ref{fig:Box}) with the interactions between the two spin fluctuations.
In the approximation where the interactions between pairs of spin fluctuations with momenta near $\mathbf{Q}_1$ and/or  $\mathbf{Q}_2$ can be treated as constants,  we use Eq. (\ref{zz_6})
 We plot the RPA form of $\chi_R(\Omega)$ in Fig. \ref{fig:Raman_full}.  We clearly see that $\operatorname{Im}\chi^0_{R}(\Omega)$ has a sharp peak at a frequency $\Omega<2\Omega_{mag}$, which is below $2\Delta$.

\section{Triangular fermion loop and symmetry channels}\label{sec:Symmetry}

So far, we have neglected the details of the Raman vertex $\Gamma_{tr}$  and thus the analysis above applies equally to all symmetry channels. We remind  that in the experiments the resonance has been observed in the $B_{1g}$ channel in the 1-Fe unit cell, but no resonance has been observed in the $B_{2g}$ or $A_{1g}$ channels. To address the origin of the difference between Raman scattering in
  different symmetry channels, we consider the symmetry properties of the function
  $\Gamma_{tr}(\mathbf{q},\nu)$, defined in (\ref{eq:Lambda}).

 First we focus on the bare Raman vertices. The original Raman vertex $\gamma_i(\mathbf{k})$ between incoming and outgoing light and incoming and outgoing fermions from band $i$ belongs to the irreducible representations of the point group $D_{4h}$. As a reminder, we point out that in the $B_{1g}$ and $B_{2g}$ representations, the function $\gamma_i (\mathbf{k})$ transforms under the point group operations as $k_x^2-k_y^2$ and $k_x k_y$, respectively, while in the $A_{1g}$ representation $\gamma_i (\mathbf{k})$
is invariant under the point group operations. We will consider each of these channels separately.

In the $B_{1g}$ channel, the Raman response is directly coupled to nematic orbital fluctuations and in the orbital basis we can define it as
\begin{equation}
	\chi_R(\Omega)= -i\int{\mathrm{d}t e^{i\Omega t} \langle T \rho_n(t) \rho_n(0) \rangle},
\label{ch_2}
\end{equation}
where the nematic "density" operator is defined as $\rho_{n}=\frac{1}{\sqrt{N}}\sum_{\mathbf{k},\sigma}\left[ a^\dagger_{\mathbf{k} \sigma}a_{\mathbf{k} \sigma}-b^\dagger_{\mathbf{k} \sigma}b_{\mathbf{k} \sigma}\right]$. In this notation, the fermion operators $a$ and $b$ correspond to the $d_{xz}$ and $d_{yz}$ orbitals, respectively.

The transformation from the orbital to the band basis in the case of the hole pockets can be approximated by [\onlinecite{vafek}]
\begin{align}\label{eq:bo_transf}
	\alpha_{\mathbf{k}\sigma}&= \cos\theta_\mathbf{k} a_{\mathbf{k}\sigma} -\sin \theta_\mathbf{k} b_{\mathbf{k}\sigma},\nonumber  \\
	\beta_{\mathbf{k}\sigma} &=  \sin\theta_\mathbf{k} a_{\mathbf{k}\sigma} + \cos \theta_\mathbf{k} b_{\mathbf{k}\sigma},
\end{align}
where $\alpha$ and $\beta$ are denote the hole bands and $\theta$ is the angle along the Fermi surface.

The contribution to the nematic density operator from fermions from the hole bands then becomes
\begin{align}
\rho_n=&\sum_\mathbf{k} (\alpha^\dagger_{\mathbf{k}\sigma} \alpha_{\mathbf{k}\sigma} -\beta^\dagger_{\mathbf{k}\sigma}\beta_{\mathbf{k}\sigma}) \cos 2\theta_\mathbf{k}\nonumber\\
 +&\sum_\mathbf{k} (\alpha^\dagger_{\mathbf{k}\sigma} \beta_{\mathbf{k}\sigma}+\beta^\dagger_{\mathbf{k}\sigma}\alpha_{\mathbf{k}\sigma})\sin 2\theta_\mathbf{k}.
\end{align}
The second term can be neglected at low energies because it couples fermions from different Fermi surfaces, which do not cross. Substituting into (\ref{ch_2}) we find that
 the vertex function
 for holes
 $\gamma_i(\mathbf{k})$ in the $B_{1g}$ channel is $\cos (2\theta_\mathbf{k})$ and has opposite signs for the two hole bands.
 For electron pockets, the situation is more simple since each electron pocket only has contributions from the $d_{zx}$ or $d_{yz}$ orbital, but not from both. The transformation from the orbital to the band basis is trivial and we find that $\gamma_i(\mathbf{k})=\pm 1$, where the plus sign is for the electrons from the pocket near $\mathbf{Q}_1=(\pi,0)$ and the minus sign for electrons near $\mathbf{Q}_2=(0,\pi)$.

For the $B_{2g}$ channel one can define a different density operator $\rho_{B2g}=\frac{1}{\sqrt{N}}\sum_{\mathbf{k},\sigma}\left[ a^\dagger_{\mathbf{k} \sigma}b_{\mathbf{k} \sigma}+b^\dagger_{\mathbf{k} \sigma}a_{\mathbf{k} \sigma}\right]$. Then using again the transformation (\ref{eq:bo_transf}) one finds that for hole bands the appropriate $B_{2g}$ symmetry factor is $\gamma_i(\mathbf{k})=\pm \sin (2\theta_\mathbf{k})$. For electron bands the symmetry factor is instead $\gamma_i(\mathbf{k})=0$ since the electron bands do not cross.

Finally, the Raman vertex in the $A_{1g}$ channel is a constant along hole or electron pockets. In general, there are two possibilities: The symmetry factor can have the same sign for the coupling of light to electrons and holes, or change sign when switching between electrons and holes. In order to consider both possibilities we define two separate functions $\gamma_{A1g}^{++}$ and $\gamma_{A1g}^{+-}$, referring to the cases with equal and opposite signs, respectively.

\begin{figure}[htb]
	\centering
		\includegraphics[width=0.4\textwidth]{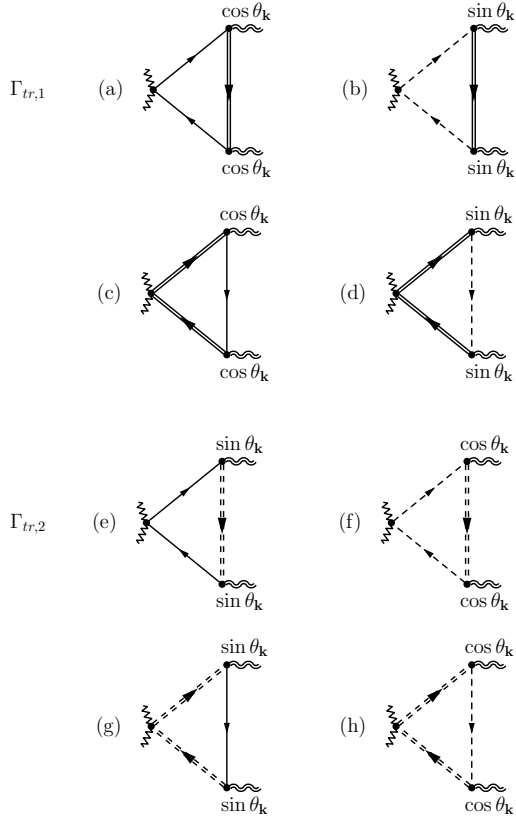}
	\caption{Contributions to the triangular fermion loop from different bands. The single solid and dashed lines represent the hole bands $\alpha$ and $\beta$, respectively, while the double solid and dashed lines represent electron bands centered at momentum $\mathbf{Q}_1$ and $\mathbf{Q}_2$, respectively. The factors of $\cos \theta$ and $\sin \theta$ arise from the transformation from the orbital to the band basis (see text).}
	\label{fig:triangle}
\end{figure}

An additional effect of the orbital to band transformation (\ref{eq:bo_transf}) is that
the factors of sine and cosine
contribute extra angular dependence to the momentum integration in $\Lambda$. This is summarized graphically in Fig. \ref{fig:triangle}, where we list the different band combinations with the appropriate signs and angular dependences. The symmetry factor $\gamma(\mathbf{k})$ for each diagram in different symmetry channels is given in Table \ref{tbl:vertex}.
The first
four diagrams [Figs. \ref{fig:triangle}(a)-(d)]are for the interaction between light and spin fluctuations with momentum near $\mathbf{Q}_1$ and the other
four are for momentum near $\mathbf{Q}_2$.  The total contribution to $\Lambda$ in each case is given by the sum of the
four diagrams. The angular dependencies listed in the figure are for a model with only intra-orbital Hubbard interaction.

\begin{table}
\centering
\begin{tabular}{|c|c|c|c|c|}
\hline
Diagram & $\gamma_{B1g}$ & $\gamma_{B2g}$ & $\gamma_{A1g}^{++}$ & $\gamma_{A1g}^{+-}$\\
\hline
Fig. \ref{fig:triangle}(a) & $\cos 2\theta$ & $\sin 2\theta$ & 1 & 1\\
\hline
Fig. \ref{fig:triangle}(b) & $-\cos 2\theta$ & $-\sin 2\theta$ & 1 & 1\\
\hline
Fig. \ref{fig:triangle}(c) & 1 & 0 & 1 & $-$1 \\
\hline
Fig. \ref{fig:triangle}(d) & 1 & 0 & 1 & $-$1 \\
\hline
Fig. \ref{fig:triangle}(e) & $\cos 2\theta$ & $\sin 2\theta$ & 1 & 1 \\
\hline
Fig. \ref{fig:triangle}(f) & $-\cos 2\theta$ & $-\sin 2\theta$ & 1 & 1 \\
\hline 
Fig. \ref{fig:triangle}(g) & $-$1 & 0 & 1 & $-$1 \\
\hline
Fig. \ref{fig:triangle}(h) & $-$1 & 0 & 1 & $-$1 \\
\hline
\end{tabular}
\caption{Symmetry factors for Raman vertices in different symmetry channels. We include two different representations of $A_{1g}$ symmetry: One where the sign changes between hole and electron pockets, and one in which it does not (see text).}
\label{tbl:vertex}
\end{table}

The result of the momentum integration is different depending on the symmetry channel. For simplicity in the evaluation of the integral, we consider identical circular Fermi surfaces in all bands and evaluate the bare triangular vertices $\Gamma^{0}_{tr}(\mathbf{q},\nu)$ in various geometries at $\mathbf{q}=0$.
  This particular value of $\mathbf{q}$ is important because the enhancement in $\chi^2 (\Omega)$ at $\Omega=\Omega_{mag}$ comes primarily from momenta near $\mathbf{q}=0$.
  Considering only the angular part of the integration, we find that
\begin{align}
	\Gamma^{0,B_{1g}}_{tr} &\propto \int \mathrm{d}\theta_{\mathbf{k}} (\cos 2\theta_{\mathbf{k}} \cos2 \theta_{\mathbf{k}} -1)\neq 0\\
	\Gamma^{0,B_{2g}}_{tr} &\propto \int \mathrm{d}\theta_{\mathbf{k}} \sin 2\theta_{\mathbf{k}} \cos2 \theta_{\mathbf{k}} = 0\\
	\Gamma^{0,{++}}_{tr}  &\propto \int \mathrm{d}\theta_{\mathbf{k}}  (1-1) =0\\
	\Gamma^{0,{+-}}_{tr}  &\propto \int \mathrm{d}\theta_{\mathbf{k}}  (1+1) \neq 0
\end{align}
where, we remind, $\Gamma^{0,{++}}_{tr}$ and $\Gamma^{0,{+-}}_{tr}$ are two different triangular vertices in $A_{1g}$ channel.
We see that $\Gamma^{0,B_{2g}}_{tr}$ and  $\Gamma^{0,{++}}_{tr}$ vanish, i.e., there is no enhancement of the Raman intensity in the $B_{2g}$ or $A_{1g}^{++}$ channels, in agreement with the data.

One fine point in the calculation is that the contribution from diagrams with electron bands at the vertex has an additional minus sign compared to diagrams with hole bands at the vertex [for example, compare diagrams Figs. \ref{fig:triangle}(c) and (d) with Figs. \ref{fig:triangle}(a) and (b)]. This is not a symmetry factor but instead comes from the opposite signs between the hole and electron band dispersions and can be obtained after performing the frequency integration in the fermion loop. In the sign-preserving $A_{1g}$ channel this leads to a complete cancellation while in the $B_{1g}$ channel it is only partial since, e.g., Figs. \ref{fig:triangle}(a) and (b) contribute a factor of $\cos 2\theta$ while Figs. \ref{fig:triangle}(c) and (d) contribute a factor of $-\frac{1}{2}\cos 2\theta$.

To summarize, we have shown so far that the bare vertices for the $B_{2g}$ and sign-preserving $A_{1g}$ channels vanish and thus cannot lead to resonances.
The bare vertices in $B_{1g}$ and sign-changing $A_{1g}$ channels are nonzero.
Note that the Raman response in the sign-changing $A_{1g}$ channel is  not screened out by long-range component of Coulomb interaction [\onlinecite{cek,blum_scr}].
The response in the sign-preserving $A_{1g}$ channel is  screened [\onlinecite{klein_1,dever,cordona}], but in our case it vanishes anyway.

The next step is to include interactions between the spin fluctuations and calculate the renormalized vertices
  $\Gamma^{B_{1g}}_{tr}$ and $\Gamma^{+-}_{tr}$.

 For this we first introduce vertices $\Gamma_{tr,1}$ and $\Gamma_{tr,2}$,
 which couple light to spin fluctuations with momentum near $\mathbf{Q}_1$ and $\mathbf{Q}_2$, respectively.
 The vertices in $B_{1g}$ and $A_{1g}$ channels are related to $\Gamma_{tr,1}$ and $\Gamma_{tr,2}$ as
 \beq
\Gamma_{tr,1}= \Gamma^{B_{1g}}_{tr} + \Gamma^{{+-}}_{tr}, ~~ \Gamma_{tr,2}= \Gamma^{{+-}}_{tr} - \Gamma^{B_{1g}}_{tr}
\eeq
 We then follow Ref. [\onlinecite{FC}] and model the interaction between spin fluctuations as given by the effective action
\begin{align}
S_{eff} &= r_0 (\Delta_1^2+\Delta_2^2) +\frac{\kappa}{2} (\Delta_1^2+\Delta_2^2)^2 \nonumber\\
	&+\frac{g}{2} (\Delta_1^2-\Delta_2^2)^2,
\end{align}
where $\Delta_1$ and $\Delta_2$ are three-component spin fluctuation fields, respectively, in which each component corresponds to a direction in real space. The $1$ and $2$ subscripts distinguish between fluctuations near $\mathbf{Q}_1$ and $\mathbf{Q}_2$. This effective action can obtained by introducing Hubbard-Stratonovich fields and then integrating out fermions. The result is [\onlinecite{FC}] that
 $\kappa >0$, but $g$ is negative, at least at small dopings.

 The bare vertices $\Gamma^0_{tr,1}$ and $\Gamma^0_{tr,2}$ are given by the sum of diagrams Figs. \ref{fig:triangle}(a)-(d) and Figs. \ref{fig:triangle}(e)-(h), respectively.  In the ladder approximation,
  the coupled equations for the renormalized  vertices $\Gamma_{tr,1}$ and $\Gamma_{tr,2}$ are given by

\begin{align}
	\Gamma_{tr,1} &= \Gamma^0_{tr,1} -(\kappa+g) \Gamma_{tr,1} \chi^2-(\kappa-g) \Gamma_{tr,2} \chi^2,\\
	\Gamma_{tr,2} &= \Gamma^0_{tr,2} -(\kappa+g) \Gamma_{tr,2} \chi^2-(\kappa-g) \Gamma_{tr,1} \chi^2,
\end{align}
where we remind that $\chi^2$ is the shorthand notation  for $\int \frac{\mathrm{d}^2\mathbf{q}\mathrm{d}\nu}{(2\pi)^3}  \chi^\mathrm{s}(\mathbf{Q}+\mathbf{q},\nu) \chi^\mathrm{s}(\mathbf{Q}+\mathbf{q},\nu+\Omega)$. In this expression we do not distinguish between $\mathbf{Q}_1$ and $\mathbf{Q}_2$ because by symmetry the result of the integration is the same.

These coupled equations can be solved in terms of $\Gamma^{B_{1g}}_{tr}$ and  $\Gamma^{{+-}}_{tr}$:
\begin{align}
	\Gamma^{{+-}}_{tr}=&\frac{\Gamma^{0,{+-}}_{tr}}{1+2\kappa \chi^2},\\
	\Gamma^{B_{1g}}_{tr}=&\frac{\Gamma^{0,B_{1g}}_{tr}}{1 + 2g\chi^2}.
\end{align}
The renormalized Raman susceptibility is then given by
\begin{align}
	\chi^{B_{1g}}_{R} = & \frac{(\Gamma^{0,B_{1g}}_{tr})^2 \chi^2}{1+2g \chi^2},\label{eq:Raman_RPA_B1g}\\
	\chi^{A1g,+-}_R = & \frac{(\Gamma^{0,{+-}}_{tr})^2\chi^2}{1+2\kappa \chi^2}.
\end{align}
 Since $\kappa >0$ and $g <0$, only the $B_{1g}$ channel has a resonance, which is consistent with the data. There
   is no resonance-type enhancement from the coupling to spin fluctuations, regardless of whether the sign-preserving or sign-changing representation is involved.
    We note in passing that there is a different enhancement of the Raman intensity in the $A_{1g}$ channel in $s^{+-}$ superconductors due to a direct process in which a light generates a ladder series of particle-hole pairs [\onlinecite{cek}].

We also note that the downward renormalization of $\Gamma^{{+-}}_{tr}$  by $1/(1 + 2\kappa \chi^2)$ also strongly reduces the temperature dependence of the $A_{1g}$ Raman intensity in the normal state.  Indeed, explicit calculation shows [\onlinecite{FC}] that $\kappa \sim 1/T^2$.
  Following the considerations of Sec. \ref{sec:T_dep}, we find that in the normal state
\begin{equation}
\chi^{A1g}_R (T) \sim \frac{(m_s)^{-2}}{T+T_1},
\end{equation}
where $T_1 \gg T_0$ because $\kappa \gg |g|$. This implies that at $T \geq T_0$,  $\chi^{A1g}_R (T)$ is nearly flat.

\section{Conclusions}\label{sec:Conclusions}
In this work we argued that the coupling of the Raman vertex to pairs of magnetic fluctuations via the AL process can explain the $1/(T-T_0)$ behavior
 of $B_{1g}$ Raman intensity in the normal state of NaFe$_{1-x}$Co$_x$As, EuFe$_2$As$_2$, SrFe$_2$As$_2$, and Ba(Fe$_{1-x}$Co$_x$)$_2$As$_2$ and the development of the resonance below $T_c$, observed in  NaFe$_{1-x}$Co$_x$As and Ba(Fe$_{1-x}$Co$_x$)$_2$As$_2$.

We considered the AL process in which light couples to a particle-hole pair, which then gets converted into a pair of spin fluctuations with momenta near $\mathbf{Q}_1 = (\pi,0)$ and $\mathbf{Q}_2 = (0,\pi)$.  We analyzed magnetically mediated Raman intensity both analytically and numerically, first at the leading two-loop order  and then included interactions between pairs of magnetic fluctuations.  We demonstrated explicitly that  the full Raman intensity in the $B_{1g}$ channel
can be viewed as the result of the
 coupling of light to Ising-nematic susceptibility via  Aslamazov-Larkin process.
 We argued that the  $1/(T-T_0)$  temperature dependence in the normal state is the combination of the temperature dependencies of the
  Aslamazov-Larkin vertex and of Ising-nematic susceptibility.  We further argued that the resonance in the $B_{1g}$ channel below $T_c$ emerges because of two effects.   One is the development of a
  pole in the fully renormalized Ising-nematic susceptibility. The pole occurs at a frequency $\Omega_{res,1} < 2 \Omega_{mag}$, where $\Omega_{mag}$
     is the minimal frequency of a dispersing spin resonance at momenta near $Q_{1,2}$ in an $s^{+-}$ superconductor.
       Another effect is that spin fluctuations generate attractive interactions
    between low-energy fermions, which constitute particle-hole bubbles with zero momentum transfer.  An attractive interaction between such fermions combined with the fact that in an $s$-wave superconductor a particle-hole bubble at zero momentum transfer is singular at $2\Delta$ gives rise to an excitonic resonance at $\Omega_{res,2} <2\Delta$.   In FeSCs $\Omega_{res,1}$ and $\Omega_{res,2}$ are not far from each other, and the observed strong peak in $B_{1g}$ Raman intensity below $T_c$ is likely the mixture of both effects.

\section{Acknowledgements}
We acknowledge with thanks conversations with  G. Blumberg, Y. Gallais, R. Hackl, R. Fernandes, M. Khodas, H. Kontani, I. Paul, A. Sacuto, J. Schmalian,  V. Thorsm\o lle, and R. Xing.  The work is supported by the DOE Grant No. DE-SC0014402.

\appendix
\section{Orbital fluctuations}\label{appendix}

\begin{figure}[htb]
	\centering
		\includegraphics[width=.49\textwidth]{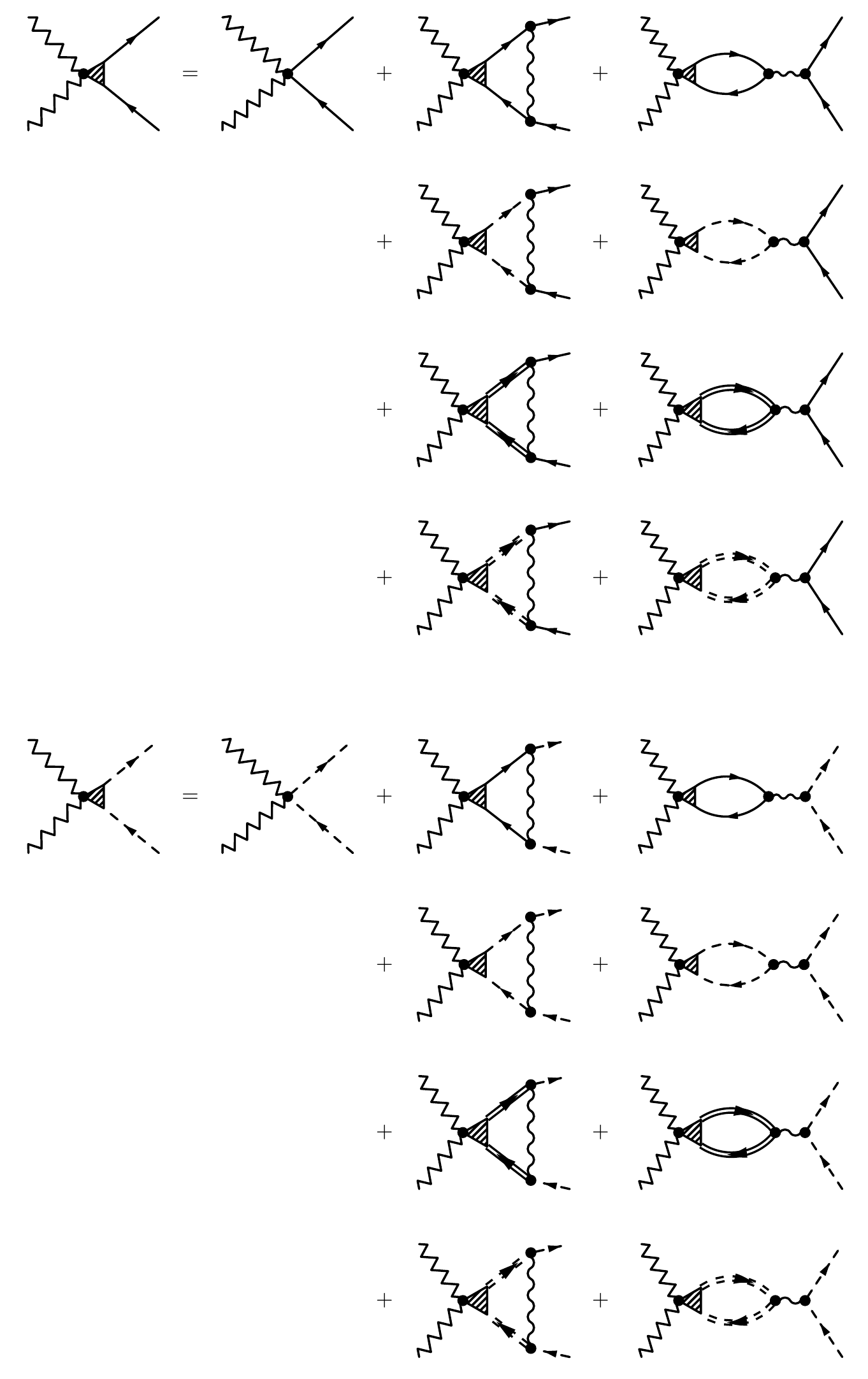}
	\caption{Vertex renormalization for hole bands $\alpha$ and $\beta$. The single solid and dashed lines represent excitations from hole bands $\alpha$ and $\beta$, respectively, and the double solid and dashed lines from electron bands $\eta$ and $\delta$, respectively.}
	\label{fig:v_renorm_h}
\end{figure}
\begin{figure}[htb]
	\centering
		\includegraphics[width=.49\textwidth]{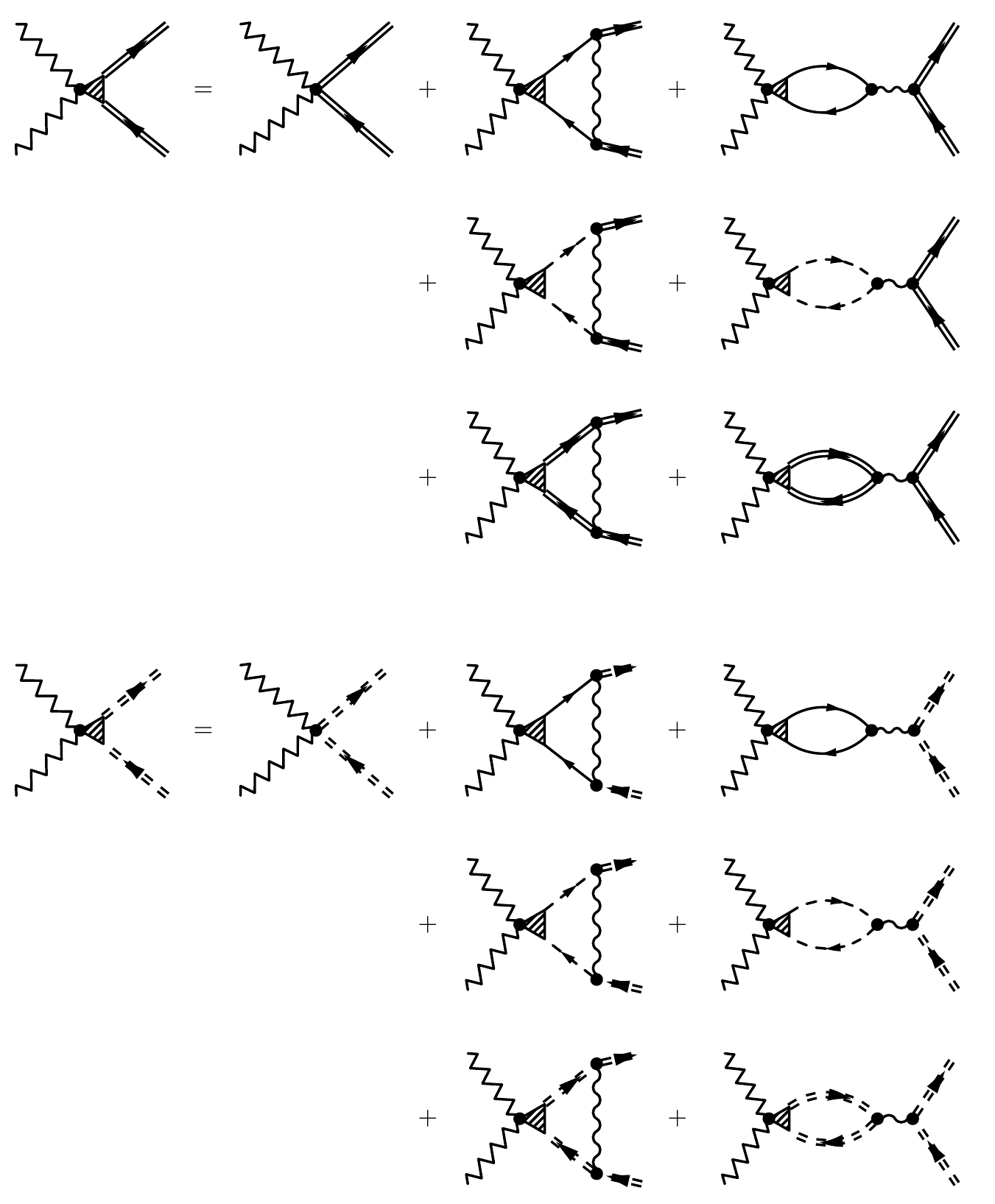}
	\caption{Vertex renormalization for electron bands $\eta$ and $\delta$. The notations are the same as in Fig.
\protect\ref{fig:v_renorm_h}.}
	\label{fig:v_renorm_e}
\end{figure}

In this Appendix we consider the coupling of the Raman response to orbital fluctuations in detail and show that the interaction in the $d$-wave channel is repulsive and cannot lead to the observed resonance.

As explained in the Introduction, the electronic structure of NaFe$_{1-x}$Co$_x$As  and Ba(Fe$_{1-x}$Co$_x$)$_2$As$_2$ consists of four bands that cross the Fermi energy. We will refer to the two hole bands centered at $(0,0)$ as $\alpha$ and $\beta$ and to the electron bands centered at $(\pi,0)$ and $(0,\pi)$ as $\eta$ and $\delta$, respectively.
We organize our analysis in the language of vertex renormalization. We start with a set of bare Raman vertices $\gamma_i(\mathbf{k})$ with both external fermion lines belonging to the $i$-th band, then dress each vertex with interactions to obtain the full vertex $\Gamma_i(\mathbf{k})$.
Here the index runs over the set of bands $\{\alpha, \beta, \eta, \delta\}$. Since we are interested in computing the Raman response in the limit of vanishingly small external momentum, we do not consider vertices with external fermion lines from two different bands since in the absence of band crossings there will be no low-energy contribution from these vertices.

For simplicity we study a model consisting of only $d_{xz}$ and $d_{yz}$ orbitals. Since the remaining $d_{xy}$ orbital has
$A_{1g}$
 symmetry it cannot directly contribute to the Raman response in the $B_{1g}$ channel. The transformation between the band and orbital basis for the hole pockets was given in (\ref{eq:bo_transf}). For the electron pockets we will simply set $\eta_{\mathbf{k}+\mathbf{Q}_1}=a_{\mathbf{k}+\mathbf{Q}_1}$ and $\delta_{\mathbf{k}+\mathbf{Q}_2}=b_{\mathbf{k}+\mathbf{Q}_2}$. Following the reasoning in Sec. \ref{sec:Symmetry}, the bare $B_{1g}$ Raman vertex is given by $\gamma_i(\mathbf{k})=\{\cos 2\theta_\mathbf{k},  -\cos 2\theta_\mathbf{k}, 1, -1\}_i$. The alternating signs reflect the difference between $d_{xz}$ and $d_{yz}$ contributions.

For our perturbative analysis, we assume that the short-range intra-orbital repulsion is the dominant interaction. Thus, the interaction Hamiltonian in the orbital basis is given by
\begin{equation}
	\mathcal{H}_I= U \sum_\mathbf{q} \left[ \rho_{xz}(\mathbf{q})\rho_{xz}(\mathbf{-q})+\rho_{yz}(\mathbf{q})\rho_{yz}(\mathbf{-q}) \right],
\end{equation}
where $\rho_{xz}(\mathbf{q})=\frac{1}{\sqrt{N}}\sum_{\mathbf{k},\sigma} a^\dagger_{\mathbf{k}+\mathbf{q} \sigma}a_{\mathbf{k} \sigma}$ and $\rho_{yz}(\mathbf{q})=\frac{1}{\sqrt{N}}\sum_{\mathbf{k},\sigma} b^\dagger_{\mathbf{k}+\mathbf{q} \sigma}b_{\mathbf{k} \sigma}$ are the density operators of the $d_{xz}$ and $d_{yz}$ orbitals, respectively.

Our diagrammatic analysis is summarized in Figs. \ref{fig:v_renorm_h} and \ref{fig:v_renorm_e}. For notational convenience we define auxiliary functions $\tilde{\Gamma}_i(\mathbf{k})$ and $\tilde{\gamma}_i(\mathbf{k})$ such that
  $\Gamma_i=\{\tilde{\Gamma}_\alpha \cos 2\theta_\mathbf{k}, \tilde{\Gamma}_\beta \cos 2\theta_\mathbf{k}, \tilde{\Gamma}_\eta, \tilde{\Gamma}_\delta\}_i$ and a similar expression for ${\gamma}_i$. The set of coupled equations for the Raman vertices can be written in matrix form as
\begin{equation}\label{eq:mvertex}
	\tilde{\mathbf{\Gamma}}= \tilde{\gamma}-\mathbf{V}\mathbf{\Pi}\tilde{\gamma},
\end{equation}
where $\mathbf{V}$ and $\mathbf{\Pi}$ are interaction and polarization matrices, respectively, given by
\begin{equation}
	\mathbf{V}= \frac{U}{2}\begin{pmatrix}
		1 & -1 & 1 & -1 \\
	 -1 & 1 & -1 & 1 \\
		1 & -1 & 2 & 0 \\
	 -1 & 1 & 0 & 2
 \end{pmatrix}
\end{equation}

\begin{equation}
	\mathbf{\Pi}= \begin{pmatrix}
		\Pi_\alpha(0,\Omega) & 0 & 0 & 0 \\
		0 & \Pi_\beta(0,\Omega) & 0 & 0 \\
		0 & 0 & \Pi_\eta(0,\Omega) & 0 \\
		0 & 0 & 0 & \Pi_\delta(0,\Omega)
 \end{pmatrix}
\end{equation}

In this notation, the polarization functions are defined as
\begin{align}
	\Pi_i(0,\Omega)=& i\int{\frac{\mathrm{d}^2 k \mathrm{d} \nu}{(2\pi)^3}} s_i(\mathbf{k})\left[ G_i(\mathbf{k},\nu+\Omega)G_i(\mathbf{k},\nu) \right. \nonumber\\
	&\quad-\left. F_i(\mathbf{k},\nu+\Omega)F_i(\mathbf{k},\nu) \right],
\end{align}
where $G_i$ and $F_i$ are normal and anomalous Green's functions
for band $i$, respectively,
 and $s_i(\mathbf{k})=\{\cos^2 (2\theta_\mathbf{k}), \cos^2 (2\theta_\mathbf{k}), 1, 1\}_i$. We note that by symmetry $\Pi_\eta (0,\Omega)=\Pi_\delta (0,\Omega)$. In this definition, the real part of each function is positive.

The solution to (\ref{eq:mvertex}) is easily obtained as $\tilde{\mathbf{\Gamma}} = (\mathbb{I}+\mathbf{V}\mathbf{\Pi})^{-1} \tilde{\gamma}$, where $\mathbb{I}$ is the identity matrix. After evaluation of the matrix multiplication we find that the full vertex is given by
\begin{equation}
	\Gamma_i=\left[1+\frac{U}{2} \sum_j \Pi_j (0,\Omega) \right]^{-1} \gamma_i
\end{equation}
and the full response function takes the simple form
\begin{equation}
	\chi_R(\Omega)=\frac{2\sum_i \Pi_i (0,\Omega)}{1+\lambda\sum_j \Pi_j (0,\Omega)},
\end{equation}
where $\lambda = U/2 >0$. This is the same formula as Eq. (2) [with $\Pi_{B_{1g}} (\Omega) = \sum_j \Pi_j (0,\Omega)$]. Obviously, for $\lambda >0$
the Raman susceptibility $\chi_R(\Omega)$ contains no poles and thus orbital fluctuations alone cannot explain the observed resonance in $\chi_R (\Omega)$.

\end{document}